\documentclass[hyper]{JHEP3}

\input{epsf}
\usepackage{epsfig}
\usepackage{amssymb}
\usepackage{amsfonts}
\usepackage{amsbsy}

\def\IC{\mathbb{C}}

\def\IZ{{\mathbb{Z}}}
\def\IR{{\mathbb{R}}}

\def\CK {{\cal K}}
\def\CN {{\cal N}}

\def\CV {{\cal V}}
\def\CW {{\cal W}}

\def\CO {{\cal O}}

\def\CK{{\cal K}}

\newcommand{\eq}[1]{Eq.~(\ref{eq:#1})}

\def\one{{\hbox{ 1\kern-.8mm l}}}

\def\bz{\bar{z}}

\def\bW{\bar{W}}
\def\bD{\bar{D}}

\def\bPi{\bar{\Pi}}

\def\bz{\bar{z}}

\def\GeV{{\rm GeV}}
\def\TeV{{\rm TeV}}

\def\ket#1{|#1\rangle}

\def\vev#1{\langle{#1}\rangle}

\def\boldclass{\bf\sf}
\def\P{{\boldclass P}}
\def\NP{{\boldclass NP}}

\def\coNP{{\boldclass{co}}\mbox{-}{\boldclass{NP}}}
\def\DP{{\boldclass DP}}

\def\PH{{\boldclass PH}}
\def\PP{{\boldclass PP}}
\def\BPP{{\boldclass BPP}}

\def\Pitwo{{\boldclass \Pi}_2}
\def\Sig#1{{\boldclass \Sigma}_#1}
\def\BQP{{\boldclass BQP}}
\def\PostBQP{{\boldclass PostBQP}}
\def\PostBPP{{\boldclass PostBPP}}
\def\PSPACE{{\boldclass PSPACE}}
\def\EXP{{\boldclass EXP}}

\def\Ppoly{{\boldclass P/poly}}
\def\BQPqpoly{{\boldclass BQP/qpoly}}
\def\PPpoly{{\boldclass PP/poly}}
\def\FP{{\boldclass FP}}
\def\FNP{{\boldclass FNP}}
\def\BPPpath{{\boldclass BPP_{path}}}

\def\itclass{\it}
\def\SAT{{\itclass SAT}}

\def\MIN{{\itclass MIN}}

\def\CC{{\itclass CC}}
\def\MINCC{{\itclass MIN-CC}}
\title{Computational complexity of the landscape I}

\author{Frederik Denef$^{1,2}$ and Michael R. Douglas$^{1,3}$\\
$^1$ NHETC and Department of Physics and Astronomy,
Rutgers University,\\
Piscataway, NJ 08855--0849, USA\\
\\
$^2$ Instituut voor Theoretische Fysica, KU Leuven, \\
Celestijnenlaan 200D, B-3001 Leuven, Belgium \\
\\
$^3$ I.H.E.S., \\
Le Bois-Marie, Bures-sur-Yvette, 91440 France\\
\\
{\tt frederik.denef@fys.kuleuven.be, mrd@physics.rutgers.edu} }

\abstract{We study the computational complexity of the
physical problem of finding vacua of string theory which agree with
data, such as the cosmological constant, and show that such problems
are typically NP hard.  In particular, we prove that
in the Bousso-Polchinski model, the problem is NP complete.
We discuss the issues this raises and the possibility that, even if
we were to find compelling evidence that some vacuum of
string theory describes our
universe, we might never be able to find that vacuum explicitly.

In a companion paper, we apply this point of view to the question of how
early cosmology might select a vacuum.}

\begin{document}

\section{Introduction}

Since about 1985 string theory has been the leading candidate for a
unified theory describing quantum gravity, the Standard Model, and all
the rest of fundamental physics.  At present there is no compelling
evidence that the theory describes our universe, as testing its
signature predictions (such as excitation states of the string)
requires far more energy than will be available in foreseeable
experiments, while quantum effects of gravity are guaranteed to be
significant only in extreme situations (the endpoint of black hole
evaporation, and the pre-inflationary era of the universe) which so
far seem unobservable.  Still, the degree of success which has been
achieved, along with the beauty of the theory and the lack of equally
successful competitors, has led many to provisionally accept it and
look for ways to get evidence for or against the theory.

Much ingenuity has been devoted to this problem, and many tests
have been proposed, which might someday provide such evidence, or else
refute the theory.  One general approach is to suggest as yet
undiscovered physics which could arise from string/M theory and not
from conventional four dimensional field theories.  Examples would be
observable consequences of the extra dimensions
\cite{Antoniadis:2005aq,Rubakov:2001kp}, or in a less dramatic vein,
cosmic strings with unusual properties \cite{Polchinski:2004ia}.

Such discoveries would be revolutionary.  On the other hand the
failure to make such a discovery would generally not be considered
evidence against string theory, as we believe there are perfectly
consistent compactifications with small extra dimensions, no cosmic
strings, and so on.  Similarly, while we might some day discover
phenomena which obviously cannot be reproduced by string theory, such
as CPT violation, at present we have no evidence pointing in this
direction.

Even if, as is presently the case, all observable physics can be well
modeled by four dimensional effective field theories, it may still be
possible to test string/M theory, by following the
strategy of classifying all consistent compactifications and checking
the predictions of each one against the data.  If none work, we have
falsified the theory, while if some work, they will make additional
concrete and testable predictions.  Conversely, if this turned out to
be impossible, even in principle, we might have to drastically
re-evaluate our approach towards fundamental physics.

While string/M theory as yet has no complete and precise definition,
making definitive statements premature, as in
\cite{Douglas:2003um}, we will
extrapolate the present evidence in an attempt to make preliminary
statements which could guide future work on these questions.  Thus,
in this work, we will consider various ingredients in the picture of
string/M theory compactification popularly referred to as the
``landscape''
\cite{Susskind:2003kw},
and try to address the following question.

Suppose the heart of the problem really is to systematically search
through candidate string/M theory vacua, and identify those which fit
present data.  This includes the Standard Model, and the generally accepted
hypothesis that the accelerated expansion of the universe is explained
by a small positive value for the cosmological constant.  This problem
can be expressed as that of reproducing a particular low energy
effective field theory (or EFT).  Suppose further that for each vacuum
(choice of compactification manifold, auxiliary information such as
bundles, branes and fluxes, and choice of vacuum expectation values of
the fields), we could compute the resulting EFT exactly, in a way
similar to the approximate computations made in present works.  Could we
then find the subset of the vacua which fit the data?  Since the
``anthropic'' solution to the cosmological constant problem requires the
existence of large numbers of vacua, $N_{vac} > 10^{120}$ or so, doing
this by brute force search is clearly infeasible.  But might there be
some better way to organize the search, a clever algorithm which 
nevertheless makes this possible?

This is a question in computational complexity theory, the
study of fundamental limitations to the tractability of well posed
computational problems.  Many of the problems which arise in
scientific computation, such as the solution of PDE's to within a
specified accuracy, are tractable, meaning that while larger problems
take longer to solve, the required time grows as a low power of the
problem size.
For example, for numerically solving a discretized
PDE, the time grows as the number of lattice sites.

Now we believe the problem at hand can be described fairly concisely,
and in this sense the problem size is small.  The data required to
uniquely specify the Standard Model is 19 real numbers, most to a few
decimal places of precision, and some discrete data (Lie group 
representations).  Although one is not known, we believe there 
exists some ``equation'' or other mathematically precise definition
of string theory, which should be fairly concise, and this is certainly
true of the approximate versions of the problem we know presently.
Thus the observation of the last paragraph would seem promising,
suggesting that given the right equations and clever algorithms,
the problem can be solved.

On the other hand, for many problems which arise naturally in
computer science, all known algorithms take a time which in the
worst case grows exponentially in the size of the problem, say the
number of variables.  Examples include the traveling salesman
problem, and the satisfiability problem, of showing that a system of
Boolean expressions is not self-contradictory (in other words there
is an assignment of truth values to variables which satisfies every
expression).  This rapid growth in complexity means that bringing
additional computing power to bear on the problem is of limited
value, and such problems often become intractable already for
moderate input size.

In the early 1970's, these observations were made precise in the
definition of complexity classes of problems 
\cite{Garey,Pap94,Rudich,Cook,Arora}.
Problems which are
solvable in polynomial time are referred to as problem class \P, and
it was shown that this definition is invariant under simple changes
of the underlying computational model (say, replacing a Turing
machine program by one written in some other programming language).
On the other hand, while no polynomial-time algorithm is known for
the satisfiability problem, a proposed solution (assignment of truth
values to variables) can be checked in polynomial time.  Such
problems are referred to as in class \NP\ (non-deterministic
polynomial). These are by no means the most difficult problems, as
in others it is not even possible to check a solution in polynomial
time, but include many intractable problems which arise in practice.
We continue our introduction to complexity theory below, and define
the ideas of \NP-complete, \NP-hard and other classes which are
widely believed to be intractable.

In this work, we observe that the class of problems which must be
solved in identifying candidate vacua in the string landscape are
\NP-hard.  The basic example is the problem of finding a vacuum with
cosmological constant of order $10^{-120}$ in natural units.  While in
suitable cases it is not hard to argue statistically that such vacua
are expected to exist, proving this by finding explicit examples may
turn out to be intractable. Indeed the intractability of similar
problems is well known in the contexts of the statistical mechanics of
spin glasses, in protein folding, and in other fields in which
landscapes of high dimension naturally appear.
We came to suspect the \NP-hardness
of the string theory problem in the course of a computer search
for flux vacua reported in the work \cite{Denef:2004dm}, and the
possibility has also been suggested independently in other works
enumerating brane constructions \cite{Blumenhagen:2004xx}.
Also, a suggestion similar to that in section 5
was made by Smolin in \cite{Smolin:2003rk},
based on the difficulty of finding 
the global minimum of a generic potential energy function.

In section 3, we prove our claim for the simplified Bousso-Polchinski
model of the landscape, and explain why we expect it for
all of the more refined models studied to date
(see \cite{Ashok:2003gk,Denef:2004ze,Denef:2004cf} and many
other physics works).
This does not necessarily imply that string theory is not testable,
just as the NP-hardness of the ground state problem for spin
glasses does not mean that the theory of spin glasses is not
testable, but clearly this point deserves serious consideration.

More interestingly perhaps from a physical point of view, these
observations lead to a paradox, analogous to one posed in the
context of protein folding \cite{Levinthal}. Namely, if it is so
difficult for us to find candidate string theory vacua, then how did
the universe do it?  As has been pointed out by various authors
(\cite{Aaronson:2005qu,Yao} and references there),
no known physical model of computation
appears to be able to solve \NP-hard problems in polynomial time. On
the other hand, according to standard cosmology, the universe
settled into its present minimum within the first few seconds after
the end of inflation, as is suggested by the correct predictions of
element abundances from models of nucleosynthesis. This would seem
far too little time for the universe to search through the
candidates in any conventional sense.

We will address this question at length in a companion paper
\cite{paperII}, but we set the stage in this paper by broadening our
discussion, beginning in section \ref{sec:relprobs}  with a survey of
other areas in physics where similar questions arise.  We then explain
in section \ref{sec:harderprob} how simple prescriptions for ``vacuum
selection principles'' arising from quantum cosmology can actually lead
to computational problems which are far more difficult than the simple
NP-hardness of fitting the data.  This will allow us to survey various
aspects of complexity theory which will be useful in the sequel,
particularly Aaronson's notion of quantum computation with postselection
\cite{Aaronson}.

In section \ref{sec:practical} we make general comments on
the somewhat more down to earth (?) question of how to test string
theory in light of these remarks, and discuss a statistical approach
to the problems they create.  We conclude by stating a speculative and
paradoxical outcome for fundamental physics which these arguments
might suggest.

We should say that while we have included some background
material with the aim of making the main points of this paper clear
for non-physicists, at various points we have assumed a fair amount
of standard particle physics background. We plan to provide a more
pedagogical account of these issues elsewhere, intended for computer
scientists and other non-physicists.

\section{The cosmological constant}
\label{sec:ccprob} \label{subsec:ccprob}

Since this is rather central to the physical motivation, let us
briefly review the cosmological constant problem for the benefit of
non-physicist readers.  More details and more references
can be found e.g.\ in
\cite{Weinberg:1988cp,Carroll:2003qq}.

Soon after Einstein proposed his theory of general relativity, he
and others began to explore cosmological models. While it seemed
natural to postulate that the universe is static and unchanging, it
soon emerged that his equations did not allow this possibility,
instead predicting that the universe would expand or contract with
time.  The reason was simply that the universe contains matter,
whose gravitational attraction leads to time dependence. To fix this
problem, Einstein added an additional term, the so-called
cosmological constant term or $\Lambda$, corresponding to an assumed
constant energy density of the vacuum.  This term can be chosen to
compensate the effect of matter so that (fine-tuned) static
solutions exist.

However, the redshift of galaxies discovered by Hubble in 1929 implies
that the universe is not static but instead expands, in a way which
was then well modeled by taking $\Lambda=0$.  Of course observations
could constrain $\Lambda=0$ only to some accuracy, and the possibility
remained open that better observations might imply that $\Lambda\ne
0$.

Now, from the point of view of quantum theory, the energy density of
the vacuum is not naturally zero, as quantum effects (vacuum
fluctuations) lead to contributions to the vacuum energy.  In some
cases, these are experimentally measurable, such as the Casimir
energy between conducting plates. Thus, there is no good reason to
forbid them in general; one might expect a non-zero $\Lambda$ of
quantum origin.

To summarize, although Einstein's original motivation for introducing
$\Lambda$ was not valid, other good motivations replaced it.  In
Weinberg's words \cite{Weinberg-phystoday}, ``Einstein's real mistake
was that he thought it was a mistake,'' and the problem of determining
the value of $\Lambda$, both observationally and theoretically, has
attracted much interest.

On the experimental side, astronomical observations long suggested 
that the average total energy density in the universe was comparable
to the total density of visible matter,
around $\rho \sim 10^{-30} {\rm g~ cm^{-3}}$.
Gradually, evidence accumulated for non-visible or ``dark'' matter
as well, in fact making up a larger fraction than the visible density.
All this was still compatible with $\Lambda=0$, however.

The situation changed in the late 1990's due to the accumulation of
new observational evidence such as type Ia supernovae, microwave
background anisotropies and dynamical matter measurements.  At
present, it is widely believed that this evidence requires our
universe to contain a ``dark energy,'' with density around
$\rho_\Lambda \sim 10^{-29} {\rm g~cm^{-3}}$ at the present epoch.

The simplest model of dark energy is a nonzero cosmological constant.
As we discuss next, obtaining such a small nonzero value for the
cosmological constant from theory is a notoriously difficult problem.
Before moving on, we should say that one can also
hypothesize alternative models for the dark energy, say with scalar
fields which are varying with time at the present epoch.  Without
going into details, all such models involve comparable small nonzero
numbers, which are typically even more difficult to explain
theoretically.  At present there is no data significantly favoring the
other possibilities, while they are more complicated to discuss, so we
will restrict attention to the cosmological constant hypothesis in the
following.  Similar considerations would apply to all the other
generally accepted models that we know about.

\subsection{Theoretical approaches to the cosmological constant problem}

We now go beyond models which describe the cosmological constant, and
discuss explaining it within a more complete theoretical framework.
Here, one long had the problem that direct computation of $\Lambda$
was not possible, because the known field theories of quantum gravity
are nonrenormalizable.  However by analogy with better understood
renormalizable field theories, one expects that in any such
computation, $\Lambda$ would be the sum of two terms.  One is a term
$\Lambda_q$ arising from quantum effects and of order the Planck
energy density, characteristic of quantum gravity. The other is a
classical, ``bare'' cosmological constant $\Lambda_0$, which in a
field theory framework is freely adjustable.

Because of this adjustable term, one had no strong theoretical
arguments favoring a particular value, but two alternatives were
generally held out.  One was that the order of magnitude of $\Lambda$
would be set by the expected order of magnitude of the quantum term
$\Lambda_q$.  Now, the fundamental scale in quantum gravity is the the
Planck energy
\footnote{We work in units with $c=\hbar=1$, and define
the Planck mass to be $M_{pl} \equiv (8 \pi G)^{-1/2} = 2.4 \times
10^{18} \,\GeV$, where $G$ is Newton's constant. $\Lambda$ will denote
the actual energy density, and not the energy density divided by
$M_{pl}^2$ as is often done in the cosmology literature. A useful online
source for energy unit conversions and constants of nature is
\cite{econv}.} 
density, $M_{pl}^4$, so one might hypothesize that
$\Lambda \sim \Lambda_q \sim M_{pl}^4$.  This is huge, of order
$10^{102}$ J/mm$^3$ $= 10^{85}$ kg/mm$^3$, i.e.\ about $10^{55}$ solar
masses in each volume unit the size of a grain of sand, and obviously
in conflict with observation.

String theory is believed to provide a well-defined theory of quantum
gravity with no free parameters, and thus one expects to be able to
compute the value of $\Lambda$.  While difficult, there have been
efforts to do this in simplified toy models.  To the extent that one
gets definite results, so far these are consistent with the assumption
that string theory is not fundamentally different from other quantum
theories for the questions at hand, in that basic concepts such as
vacuum energy, quantum corrections, effective potential and so forth
have meaning, are calculable from a microscopic definition in a
similar way to renormalizable quantum field theories, and behave much
as they do there and in semiclassical quantum gravity.  One can
certainly question this idea \cite{Banks:2004xh}, but since it enters
at a very early stage into all present attempts to make detailed
contact between string theory and the real world, major revisions of
our understanding at this level would require restarting the entire
theoretical discussion from scratch.  We see no compelling reason to
do this, and proceed to follow this widely held assumption.

In some quantum theories, especially supersymmetric theories, the
vacuum energy is far smaller than naive expectations, due to
cancellations.  Now there is no reason to expect this for the well
established Standard Model, but it might apply to a hypothetical
extension of the Standard Model which postulates new fields and new
contributions to the vacuum energy at some new fundamental energy
scale $M_f$, which would set the scale of $\Lambda_q$.
Experimental bounds on new particles and forces
require roughly $M_f \ge 1 \,\, \TeV$, and even if this bound were
saturated, the vacuum energy would still be one Earth mass per mm$^3$.

We do not and could not live in such a universe. If $\Lambda = M_{pl}^4$,
the universe would have a Planck scale curvature radius and expand
exponentially with a Planck scale time constant. If $\Lambda = (1 \,\,
\TeV)^4$, it would still have a sub-millimeter scale curvature radius
and inflate exponentially on a time scale of less than a
picosecond. For negative cosmological constants of this order the
universe would re-collapse into a Big Crunch on these ultrashort time
scales.  Thus, the simple fact of our own existence requires $\Lambda$
to be extremely small.

This led many to the second, alternative hypothesis, which was that
$\Lambda$ should be exactly zero, in other words the adjustable term
$\Lambda_0=-\Lambda_q$, for some deep theoretical reason.  Now in
analogous problems, there often are arguments which favor the
parameter value zero.  For example, if a nonzero value for a parameter
$\alpha$ breaks a symmetry, one can argue that quantum corrections
will themselves be proportional to $\alpha$, making $\alpha=0$ a
self-consistent choice.  While no fully convincing argument of this
type for $\Lambda=0$
has ever been found, the simplicity of this hypothesis
makes it hard to ignore.
Thus it was that theorists found themselves in a way repeating
``Einstein's mistake'' in reverse, unable to predict with any confidence
that $\Lambda\ne 0$ was a serious possibility until observation made
it apparent.

Now there was one pre-1990's theoretical idea which did lead to such
small values in a fairly natural way.  This was that the fundamental
theory should contain a large number $N_{vac}$ of vacuum
configurations, realizing different effective laws of physics, and in
particular with different values of $\Lambda$.  As we will see later,
this is easy to achieve in field theory, in many ways, and appears to
be true in string theory.  Given an appropriate distribution of
$\Lambda$ values, it then becomes statistically likely that a vacuum
with the small observed value of $\Lambda$ exists.

To illustrate, suppose the number distribution of $\Lambda$ among vacua were
roughly uniform, meaning that the number of vacua with $\Lambda$
in an interval $\Lambda\in(a,b)$ were roughly 
$$
N_{vac}(a \le \Lambda \le b) \sim N_{vac}\frac{b-a}{2 M_{pl}^4} ,
$$
for any $|a|,|b|\le M_{pl}^4$.  If so, the claim that there exists a vacuum 
with $\Lambda \sim M_{pl}^4/N_{vac}$ becomes statistically likely, and this
might be considered sufficient evidence to take 
the claim that such a theory can solve the cosmological constant problem
seriously.
Of course, there are clearly pitfalls to guard against here, such as the
possibility that the distribution has unusual structure near $\Lambda=0$,
correlations with other observables and so forth, but 
keeping these in mind, let us proceed.

This argument does not yet explain why we find ourselves in a vacuum
with a small value of $\Lambda$.  One might expect such a question to
be explained by the dynamics of early cosmology, and thus try to
identify a dynamical mechanism which leads with high probability to a
vacuum with an extremely low or perhaps even the lowest positive
$\Lambda$.  We will discuss this idea below and in \cite{paperII}, but
for now let us just say that this appears problematic.

A different approach is to claim that all of the possible vacua
``exist'' in some sense, and that at least part of the structure of
the vacuum we observe is simply environmentally selected: we can only
find ourselves in a vacuum where the conditions are such that
something like an observer asking these questions can exist.  This
general idea is known as the ``anthropic principle'' \cite{Barrow}.
Since as we mentioned, the simple fact of our own existence
requires $\Lambda$ to be extremely small, this principle would seem
highly relevant here.

Various objections have been raised to the anthropic principle.  While
at first it may seem tautological, it clearly does have content in the
context of a physical theory which predicts many possible candidate
laws and structures for our universe, as it often provides a very
simple answer to the question ``why do we not find ourselves in vacuum
X.''  There are more serious objections.  It is unimaginably difficult
to characterize the most general conditions which might allow for
observers who can ask the questions we are addressing.  Even
restricting attention to simple necessary conditions, analyzing their
dependence on the many fundamental parameters is complicated.
Finally, if there are many ``anthropically allowed'' vacua, it does
not lead to any clear preference among them.  But keeping these
caveats in mind, the principle has led to interesting claims.

The original anthropic argument bearing on the cosmological constant 
took as the necessary condition the requirement that some sort of structure
such as galaxies, could form from an initially smooth post Big Bang
universe by gravitational clumping of matter. This puts a bound on
the cosmological constant because if $\Lambda$ is
significantly bigger than the matter density at the time when
gravitational clumping starts, the acceleration of expansion driven
by $\Lambda$ will outpace the clumping process, and the universe
ends up as a dilute, cold gas. Assuming all other parameters fixed,
Weinberg \cite{Weinberg:1987dv} computed in 1987 that this requires
$\Lambda < 400 \rho_0$, where $\rho_0$ is the present matter
density. A (negative) lower bound is obtained by requiring the
universe to survive for long enough to allow for some form of life
to evolve, before re-collapsing into a Big Crunch. Again assuming
all other parameters fixed, this gives a bound $\Lambda > - \rho_0$.
Together this gives the allowed window
\begin{equation} \label{weinbergwindow}
 - 10^{-120} M_p^4 < \Lambda < 10^{-118} M_p^4.
\end{equation}
Although there are a number of assumptions that went into this
computation, most notably fixing the amplitude of primordial density
fluctuations, and although the window is still two orders of
magnitude wider than the observed value of $\Lambda$, this is by far
the most successful (and simple) computation of $\Lambda$ produced in
any theoretical framework to date.  Indeed, it might be regarded
as a prediction, as it came well before the evidence.

While this argument is clearly important, as we discussed we now have
direct evidence for non-zero dark energy, and thus part of ``fitting
the data'' is to reproduce this fact.  Despite many attempts, at
present the only theoretical approach in which we can convincingly
argue that this can be done is the statistical argument we discussed.

\subsection{Landscape models}

A ``vacuum'' is a candidate ground state of a physical theory.  It
should be either time independent, {\it i.e.} stable, or extremely long lived
compared to the physical processes under consideration,
{\it i.e.} metastable.
Since we seek a theory
which can describe all physics through all observed time, its
average lifetime
must far exceed the current age of the universe, of
order $10^{10}$ years.

In well understood theories, to a good approximation stability is
determined by classical considerations involving an energy functional
called the effective potential (part of the EFT mentioned in the introduction).
Both stability and metastability require that the energy increases
under any small variations of the configuration.  In quantum theory,
one can also have instability caused by tunneling events to lower
energy configurations.  Thus, acceptable metastability requires
that the barriers to tunneling are so high that tunneling rates are
negligible on scales far exceeding the current age of the universe.

A simple model for the vacuum energy and its dependence on the
configuration is to describe the configuration as a ``scalar field,'' a
map $\phi$ from space-time into some manifold $\cal C$.  The vacuum energy 
functional $E$ is then determined by a real valued function $V$ on $\cal C$,
the potential.  It is given by an integral over all space, which if the
derivatives of the fields are small has an expansion
$$
E = \int d^3x \sqrt{g}~ V(\phi(x)) +
 \CO(\left(\frac{\partial\phi}{\partial x}\right)^2) .
$$
In this case, a vacuum is a constant field configuration $\phi=\phi_0$
with $\partial\phi/\partial x = 0$, which is a local minimum,
{\it i.e.} a critical point with positive definite Hessian,
$$
\frac{\partial V}{\partial \phi} = 0 ; \qquad
\frac{\partial^2 V}{\partial \phi\partial\phi} > 0 .
$$
The value at the minimum $\Lambda=V(\phi_0)$ is the energy of the vacuum,
in other words the cosmological constant in that vacuum.

Such a model already suffices to realize the scenario we described,
in which there are many vacua realizing widely differing values of
$\Lambda$.  The simplest models of this type, which were also the
first to be proposed \cite{Abbott},
simply take $\cal C=\IR$ and a ``washboard'' potential such as
\begin{equation} \label{eq:simpleV}
V(\phi) = a \phi - b \cos 2\pi\phi .
\end{equation}
For $a<<b$, this potential has many minima, with equally spaced
cosmological constants $\Lambda=a n - b$ for all $n\in\IZ$.  Thus we simply
need $a\sim 10^{-120} M_{pl}^4$ for some vacuum to realize the observed
$\Lambda$.  There are many variations on this construction, which solve
the problem with regular potentials depending on a few fields.

Instead of continuously adjustable fields $\phi$ and a potential with
isolated minima, another possibility is to find models with discrete
choices.  One systematic discrete choice, which will be exploited in
the Bousso-Polchinski model below, is that of ``flux.'' This is
a postulated magnetic field (or generalized ``$p$-form magnetic field'')
in the extra dimensions.  
As in electrodynamics, such a magnetic field will contribute to the
potential energy as $V \sim B^2$.  But whereas ordinary magnetic field breaks
rotational symmetry, this one only breaks symmetries in the extra
dimensions, so such a configuration is still a ``vacuum.''  One might
ask why it is stable, as a localized configuration of
ordinary  magnetic field is not.  This is because of
Dirac's quantization condition for magnetic charge.
On a topologically nontrivial manifold $M$, this generalizes to the
statement that, after expressing the magnetic field in appropriate units,
its integral over any homology cycle in $M$ is an integer, and
thus the field cannot vary continuously or decay (classically).

While this is different microscopically from
a scalar field, its effect
on observed physics is not very different, and roughly corresponds
to taking $\phi\in\IZ^n$ in a potential like the ones we just 
postulated.\footnote{One might think that the two types of model differ
in whether vacua with different values of $\phi$ are
connected by physical time evolution, as the flux cannot change
classically.  However, in both cases $\phi$
can change via tunneling events.}
One could again postulate the potential \eq{simpleV} to get a
candidate solution to the cosmological constant problem, or even
derive it from a microscopic model as in 
\cite{Brown:1987dd,Brown:1988kg}.

The models we just discussed have the serious flaw that the problem of
obtaining the small coefficient $a\sim 10^{-120} M_{pl}^4$ in \eq{simpleV}
is just as hard as solving the original cosmological constant problem.
A way around this, suggested in \cite{Banks:1991mb}, is to take
${\cal C}=\IR^2$ or $\IZ^2$, with coordinates $\phi^1$, $\phi^2$, and
\begin{equation} \label{eq:simpleVtwo}
V(\phi) = a_1 \phi^1 - b_1 \cos 2\pi\phi^1
        + a_2 \phi^2 - b_2 \cos 2\pi\phi^2 + b_1 + b_2
\end{equation}
If the ratio $a_1/a_2$ is irrational, then the set of possible
cosmological constants $\Lambda=a_1 n_1+a_2 n_2$ will be dense in
$\IR$, without the need to postulate small numbers.  
To get a finite but large set, one can bound the range of
$\phi^1$ and $\phi^2$.

While these models would not lead to the complexity issues we are
about to discuss, they are also drastic oversimplifications of the
real models which arise from string theory compactification.  In
practice, the potential is never as simple as \eq{simpleV} or
\eq{simpleVtwo}; indeed the problem of finding any controlled
description of the potential which is valid over the large range of
$\phi$ values required in either of these models
is difficult.  While the discussion rapidly
gets technical, in the better understood models
arising from string theory, such as
\cite{Giddings:2001yu,Ashok:2003gk,Denef:2004cf,Denef:2004ze}, one finds that
in models with few fields, the required large number of 
quasi-realistic minima is not present.

But this is not the only way we could imagine finding a large set 
(or ``discretuum'' \cite{Bousso:2000xa})) of values of $\Lambda$.
Another general possibility
is for $\cal C$ to be a space of high
dimension, with $V$ a fairly generic function on $\cal C$.
In string theory compactification, $\dim\cal C$ is typically determined
by a topological invariant of the compactification manifold, say a
Betti number, and can take values ranging up to $1000$ or more.
As is familiar in the study of optimization
problems, and as we are about to discuss
at great length, generic functions on high dimensional spaces typically
have a number of minima which grows exponentially in the dimension,
and realize many different values at the local minima.  

Potentials of this type,
depending on many (continuous or discrete) variables,
appear naturally in many areas of physics and mathematical
biology, and are often referred to as ``landscapes.''\footnote{
The term first appeared in the phrase ``fitness landscape'' in
evolutionary biology \cite{Wright}.  While we are not historians,
from perusal of reviews such as \cite{Wales,Mezard},
it appears that systematic consideration of energy landscapes in physics
and physical chemistry began in the 1970's, with the term ``landscape''
coming into common though occasional use in the mid-1980's.}
As we discuss shortly, they also
appear naturally in string theory, which has led to the name
``string theory landscape'' \cite{Susskind:2003kw}. 

It should be clear that the idea of a landscape is \emph{a priori}
completely independent of the idea of anthropic or environmental
selection principles: even in the presence of a landscape, one could
in principle imagine a strong dynamical selection mechanism producing
the apparent fine-tunings we observe.  Thus we will try to carefully
distinguish these ideas in the following.

%\vskip 0.5in

\section{Computational complexity of finding flux vacua}
\label{sec:fluxNP}

We begin by reviewing the Bousso-Polchinski (BP) model
\cite{Bousso:2000xa}, to set out the problem. We then turn to issues
of computational complexity in this context. As a warm-up example we
consider a toy model of Arkani-Hamed, Dimopoulos and Kachru
\cite{Arkani-Hamed:2005yv}, for which the analysis is very easy. We
then proceed to the detailed analysis of the complexity of the BP
problem, and briefly discuss related problems and more sophisticated
models of the string theory landscape.

\subsection{The Bousso-Polchinski model}

Stripped down to its essence, the Bousso-Polchinski model
consists of a set of $K$ quantized fluxes given by a vector $N \in
\IZ^K$, giving rise to a vacuum energy
\begin{equation} \label{eq:BPmodeldef}
 \Lambda = \Lambda_0 + \sum_{ij} g_{ij} N^i N^j.
\end{equation}
Here $\Lambda_0$ is some bare negative cosmological constant (which
can be thought of as coming from orientifold, curvature and other
flux-independent contributions), and $g_{ij}$ some positive definite
metric. The dimension $K$ is determined by topological invariants
of the compactification manifold; such as a Betti number; in
examples this can range up to $1000$ or so.
In general $g_{ij}$ and $\Lambda_0$ will depend on many
scalar fields (moduli), which in turn are stabilized at minima of
the vacuum energy function. To avoid having to deal with such a
complicated coupled system, the BP model simply assumes the moduli
to be frozen, and $g_{ij}$ and $\Lambda_0$ to be given constants.
The fully coupled system as arising in actual string
compactifications was analyzed in
\cite{Ashok:2003gk,Denef:2004cf,Denef:2004ze}, confirming the
qualitative picture emerging from the BP model.

\EPSFIGURE{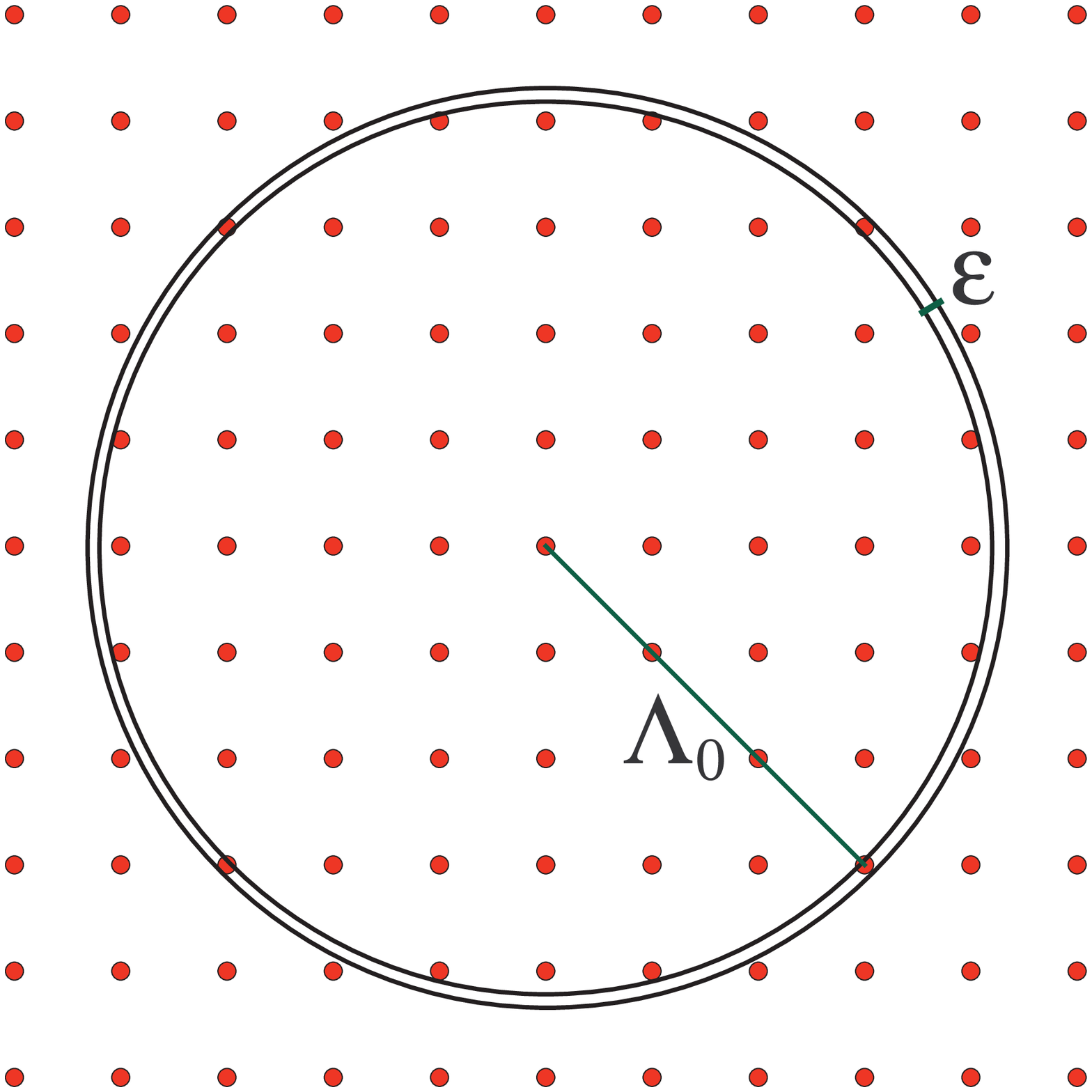,height=6cm}{Scanning the cosmological constant in
the Bousso-Polchinski model \label{bp}}

A crucial observation of \cite{Bousso:2000xa} was that for
sufficiently large $K$ and $|\Lambda_0|$, the cosmological constant
$\Lambda$ gets sufficiently finely scanned near zero to allow for
extremely small values, including numbers of the order of the
presently observed cosmological constant $\Lambda \sim 10^{-120}
M_p^4$. The idea, sketched for $K=2$ in figure \ref{bp}, is that for
large $K$ and $|\Lambda_0|$, even a very thin shell of radius
$\sqrt{-\Lambda_0}$ will contain some lattice points.

The basic estimate is as follows. A ball defined by $g_{ij} N^i N^j
\leq L$ has Euclidean volume
\begin{equation}
 {\rm Vol}(L) = \frac{(\pi L)^{K/2}}{(K/2)!} \frac{1}{\sqrt{\det g}}
\end{equation}
in $N$-space. For ${\rm Vol}(L) \gg 1$, this is a good estimate for
the number of lattice points contained in the ball. Similarly, a
thin shell of radius squared $|\Lambda_0|$ and width $\epsilon$ has
a volume
\begin{equation} \label{Nvacestimate}
 \delta {\rm Vol} \approx \frac{K}{2 |\Lambda_0|} \,  {\rm
 Vol}(|\Lambda_0|) \, \epsilon
\end{equation}
and hence when this is much bigger than 1, we expect about $\delta
{\rm Vol}$ vacua (labeled by different values of $N$) with a
cosmological constant of order $\epsilon$.

Suppose however that we want to do better than these estimates and
find out exactly which vacua have cosmological constant $\Lambda \in
(0,\epsilon)$, or at least find out with certainty whether or not
there is one in this interval.\footnote{Of course, to make this a
meaningful question, we need to know all given quantities to a
sufficiently high precision. We assume this for now but will discuss
this issue in more detail further on.} A related problem is to find
the vacuum with the smallest positive cosmological constant. The
first problem would be a maximally simplified model of finding
string vacua in agreement with observational data in physical
parameter space. The second kind of problem would naturally arise
for instance if we grant an {\it a priori} probability distribution
$P(\Lambda) \sim \exp(c/\Lambda)$ on the landscape of vacua, since
such a distribution strongly peaks on the smallest positive value of
$\Lambda$. This particular choice of $P(\Lambda)$ can be motivated
in various cosmological frameworks; it will be discussed in detail
in later sections.

Clearly, none of these problems are easy. But before we go on and
quantify just \emph{how} hard they are, we consider an even simpler
model, for which the complexity analysis is more straightforward.

\subsection{A toy landscape model} \label{subsec:toymodel}

In \cite{Arkani-Hamed:2005yv}, Arkani-Hamed, Dimopoulos and Kachru
(ADK) proposed a field theory toy model for the landscape consisting
of $N$ scalars $\phi^i$ and a potential
\begin{equation}
 V(\phi) = \sum_i V_i(\phi^i)
\end{equation}
where each $V_i$ has two minima, $V_i^- < V_i^+$. This model has
$2^N$ vacua, with vacuum energies
\begin{equation} \label{eq:ADKenergies}
 \Lambda_m = \sum_i m_i (V_i^+ - V_i^-) + V_i^- = \sum_i m_i \Delta V_i \,
 + \, V_{\rm min}
\end{equation}
where $m_i \in \{ 0,1 \}$. The distribution of vacuum energies was
studied in \cite{Arkani-Hamed:2005yv} in the large $N$
approximation, in which the central limit theorem can be used to
argue that this distribution will be a Gaussian.

Suppose however that we want to find the precise smallest positive
value of $\Lambda$ in this ensemble, or want to find out with
certainty if there exists a vacuum with $\Lambda$ within the range
$(\Lambda_0,\Lambda_0+\epsilon)$. The most naive algorithm to solve
this problem would be to simply loop through all vacua. This would
take $2^N$ steps, which becomes exponentially quickly infeasible for
large $N$; for $N=50$, scanning one billion instances per second,
this would take about two weeks. For $N=400$, the minimal value to
(naively at least) expect vacua with $\Lambda$ of order $10^{120}$,
it would take $10^{104}$ years.

\subsection{A brief introduction to complexity theory}

\label{subsec:introcompl}

Can the preceding algorithm be improved to give a solution in some
time polynomial in $N$ rather than exponential in $N$? Questions
like this are the subject of computational complexity theory, to
which we now give a brief introduction.

\EPSFIGURE{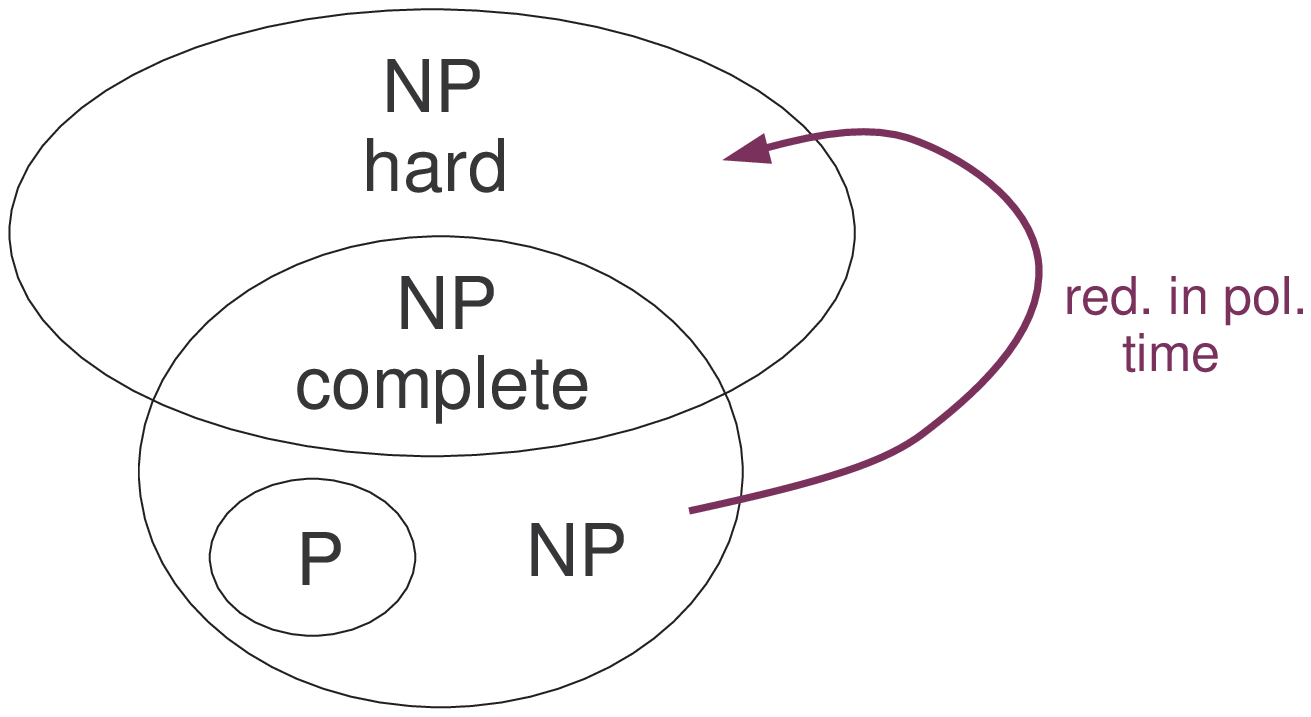,height=4cm}{Basic complexity classes
  \label{nphard}}

The following basic complexity classes are relevant for the problem
at hand:\footnote{See \cite{Pap94,Rudich,Arora} for a detailed
explanation, and \cite{complexityzoo} for a huge up-to-date list of
complexity classes. For a more elementary online introduction, see
\cite{wikipedia:complexity}.}
\begin{itemize}

 \item {\P} ~(Polynomial time): Decision (yes/no) problems solvable by an
 ordinary computer (or more formally a deterministic Turing
 machine) in a number of steps polynomial in the input size. For
 example the decision problem ``Is $N = N_1 \cdot N_2$?'' is in \P, since
 the time needed to multiply two integers is polynomial in the number
 of digits. A highly nontrivial example of a polynomial time problem
 is primality testing \cite{AKS02}.

 \item {\NP} ~(Non-deterministic Polynomial time): Decision problems for
 which a candidate solution can be verified in polynomial time. More
 precisely, if the answer is yes, there should exist a proof of this
 verifiable in polynomial time. Typically an \NP\ problem is of
 the form ``Does there exist an $x$ such that $p(x)$ is true?'',
 where ``$p(x)$ is true'' can be verified in polynomial time for any
 given $x$. The problem of finding out whether or not there exists a
 vacuum in the ADK toy landscape with energy in a specified range is
 in \NP, since a candidate example can be checked in time
 polynomial in the input size.\footnote{We are assuming here that the energies
 $V_i^\pm$ are given up to a fixed precision and represented as
 rational numbers. We will discuss precision issues further on.}

 Equivalently, these are the problems that can be
 solved in polynomial time on a \emph{non-deterministic} Turing
 machine. Roughly speaking, the difference between a
 deterministic Turing machine and a non-deterministic one is that the
 former works like a standard computer, executing instructions sequentially in
 one chain, whereas the latter is allowed to branch off in a
 number of ``copies'' at each step, with each copy executing
 a different operation at that step. Hence, rather than a single
 computation path, we now have a computation tree. If the answer is yes,
 at least one of the computation paths should accept. If the answer is no,
 all computation paths should reject. For the toy landscape problem,
 the branches at level $i$ of the tree could be taken to correspond to picking
 $m_i=0$ resp.\ $m_i=1$. Alternatively, one can think of the branching
 as flipping a coin.
 \NP\ problems are then problems for which a polynomial time
 randomized algorithm exists that always outputs no when the answer
 is no, and at least in one possible run outputs yes when the answer is
 yes.

 Obviously ${\P} \subseteq {\NP}$.

 \item {\NP-hard}: Loosely speaking, this is the class of problems at
least as hard as \emph{any} \NP\ problem. More precisely this means
that any \NP\ problem can be reduced to it in polynomial time, in a sense
we will explain in an example shortly.

 \item {\bf \NP-complete}: An \NP-hard problem that is in \NP\ itself is
 called \NP-complete. In this sense they are the hardest problems in
 the class \NP. It is quite remarkable that natural
 \NP-complete problems exist at
 all, but by now a large collection of such problems is
 known. One simple example is the subset-sum problem:
 given a finite list of integers and a target integer $t$, 
does there exist a subset that sums up to $t$?
\end{itemize}
By definition, if \emph{any} \NP-hard problem can be solved in
polynomial time, \emph{all} \NP\ problems can be solved in polynomial
time, and therefore we would have \P=\NP. It is widely believed that
this is not the case, although there is no proof, and the \P\ versus
\NP\ problem is considered to be the biggest open question in
theoretical computer science. So big in fact that it is one of the
seven Clay Millennium Prize problems \cite{clayprize}.

While we will not need complete and precise definitions of these
concepts for our discussion, a point worth keeping in mind is the
difference between decision problems, which take an input and
produce a yes or no output, and problems which produce more general
outputs. The classes we just described refer to decision problems.
While similar definitions can be made for other problems (say for
computing a function, we have \FP, \FNP\ and so on), to simplify our
discussion we will stick to decision problems.

Thus, we pose the ADK toy landscape problem: does the set of vacuum
energies $\Lambda_m$ obtained from \eq{ADKenergies}, contain a
vacuum in the range $(\Lambda_0,\Lambda_0+\epsilon)$ ?

The bad news is that this problem is already
\NP-complete. It is a special case of the classic \emph{knapsack
problem} (and closely related to the subset sum problem): given a
list of items with cost $c_i$ and value $v_i$, can a subset of items
be selected such that the total value exceeds a given $v_{\rm min}$
but the total cost remains below $c_{\rm max}$? This problem is well
known to be \NP-complete \cite{karpNPC,Garey} and reduces to our
landscape problem by taking
$$
 v_i = c_i = \Delta V_i, \quad v_{\rm min} = \Lambda_0 - V_{\rm min},
 \quad c_{\rm max}=\Lambda_0 - V_{\rm min} + \epsilon.
$$
This means that unless \NP\ is the same as \P, there is no polynomial
time algorithm to solve our problem exactly for all possible
instances.

While the proof of \NP-completeness of the knapsack problem or the
closely related subset sum problem is given in textbooks
\cite{Garey,Pap94}, let us briefly summarize the approach. Suppose we
want to prove \NP-completeness of subset sum.  First, we choose a
known \NP-complete problem.  Then, we show that for any instance of
that problem, we can construct a subset sum problem whose solution
could be translated back into a solution of the original problem.
This is referred to as a ``reduction'' of our original problem to
subset sum, and the last point to check is that this reduction must be
doable in a polynomial number of elementary operations.

A convenient choice for the known \NP-complete problem
is the satisfiability or \SAT\ problem. In this
problem, one is given a list of $N$ Boolean variables, each taking
the value true or false, and a list of clauses, each a Boolean OR in
which a subset of the variables appear, either literally or negated.
For example, labelling the variables $x_i$, a sample clause would be
$$
x_2 \vee \bar{x_7} \vee x_{22}
$$
where the bar represents NOT, and $\vee$ represents OR.  The problem
is to find an assignment of truth values to variables which
satisfies all of the clauses.  Since a proposed assignment can be
checked in time proportional to the number of clauses, this problem
is in \NP.  A closely related variant is the
$k$-\SAT\ problem, in which a definite number $k$ of variables
appear in each clause. One can show that any \SAT\ problem can be
(polynomially) mapped into a $3$-\SAT\ problem, by introducing new
variables and subdividing clauses.

The \SAT\ problem was shown to be \NP-complete \cite{Cook} as
follows. We need to argue that, given a Turing machine program which
verifies a solution to some problem in polynomial time, we can
translate this verification into a \SAT\ problem.  This is done by
considering a ``tableau'' which is simply the runtime history of a
Turing machine computation, step by step.  Now, any valid history
will satisfy a set of logical relations between subsequent steps,
whose simultaneous truth implies that some particular run of the
program results in the machine deciding that the original problem
has a solution.  These relations give us a \SAT\ problem with a
polynomial number of clauses, whose solution amounts to the
verification of a solution to our original problem.  By feeding this
to a hypothetical polynomial time \SAT\ solver, we could solve the
original decision problem in polynomial time.

The reduction of \SAT\ (or 3-\SAT) to subset sum is analogous
although of course the details are different.  Given a set of
Boolean clauses, one needs to construct a set of numbers which will
sum to a given number only if the 3-\SAT\ problem has a satisfying
assignment.  By the trick of using ``base $b$'' numbers of the form
$N=\sum c_k b^k$, one can express the condition that many different
(small) $c_k$'s sum to specified numbers simultaneously, so we just
need to encode the conditions that each variable is either true or
false, and each clause is true, into a set of subset conditions on
independent $c_k$'s. In appendix \ref{app:NPCss}, we give a detailed
proof based on this idea.

%To illustrate, let us introduce a pair of numbers $N_{2i}$ and
%$N_{2i+1}$ for each variable $x_i$, each with $c_i=1$, and require
%the sum to have $c_i=1$.  The subset sum condition on $c_i$ then
%forces a single number from $\{N_{2i},N_{2i+1}\}$ to appear in the
%subset, while conditions on other $c_j$ coefficients can then be
%used to enforce the truth of the clauses.  The reader may enjoy
%finding such an encoding to complete the argument.

The upshot is that finding a general polynomial time subset sum
solver would imply $\P=\NP$, which is generally believed to be
false, and thus we believe that the answer to the question we began
the subsection with is ``no.''

\subsection{Approximate algorithms, and physical approaches}

The previous argument, while standard in computer science,
is a bit intricate and quite different from the
usual physical discussions of such problems.  Now such phenomena
are already known in other branches of physics, and we discuss these
connections in section \ref{sec:relprobs}.
But let us first discuss some possible ways
around it, by relaxing the rather strong requirements we set out, or
adopting more physical approaches.

First, while the argument we just gave shows clearly that the
complexity of the problem grows with the number of fluxes, it does not
directly refer to the size or accuracy of the energies we are considering.
Indeed, in practice, we will usually
not know the exact values of the $V_i^\pm$, even if they are
computable in principle from a fundamental theory.  For example, various
quantum corrections may be quite hard to compute.\footnote{These
quantum corrections will typically also spoil the simple direct form
of the potential, but we will ignore this for the sake of the
argument.}

Thus, let us assume we know the values of the $V_i^\pm$ to a given
accuracy of order $\delta$. The problem we posed is then only
sensible of course if $\delta < \epsilon/N$.  Now for \emph{fixed}
values of $\delta$ and $\Delta V_i$, the solution can be found
(within the maximal precision determined by $\delta$) in time
polynomial in $N$. In appendix \ref{app:knapsackalg}, we give a
simple algorithm solving the problem in $O(N (\sum_i \Delta V_i) /
\delta)$ time. If an order $\epsilon$ error in the target energy
range is allowed, we can assume $\delta = \epsilon/N$, so the
required time is of order $N^2 (\sum_i \Delta V_i) / \epsilon$.
Hence if $\epsilon$ (or $\delta$) is not much smaller than the
$\Delta V_i$, the problem becomes effectively tractable. This is an
example of an \NP-hard problem with a ``fully polynomial time
approximation scheme'' (FPTAS). By no means are all \NP-hard
problems approximable in this strong sense, see e.g.\
\cite{NPOprobs,KleinYoung}.

The catch is that if we want this toy landscape
problem to model the problem of finding out whether there exist
string vacua with cosmological constant of order $10^{-120} M_p^4$,
and the typical value of $\Delta V_i$ is Planck scale, we should
take $\sum_i \Delta V_i / \epsilon \sim 10^{120}$.  Then, the algorithm
of appendix \ref{app:knapsackalg} becomes effectively useless.
This illustrates that it is also the accuracy to with which we must solve
the problem, as well as the large number of fluxes, which makes it difficult

We can contemplate different, more ``physical'' algorithms to find
vacua with minimal or small cosmological constant. For example, we
could just start off a particle high up a hill of $V(\phi)$ and try
to let it roll down to a vacuum with cosmological constant in
$(0,\epsilon)$. Clearly for small $\epsilon$ this is not going to
work: there are so many local minima at values of $\Lambda$ well
outside $(0,\epsilon)$ that we would quickly get stuck, and would
almost never end up in the target interval. A similar conclusion
holds for a discretized version where we stick to the actual vacua
and define local jumps as flipping one $m_i$ between $0$ and $1$. We
could add thermal noise to the motion of the particle such that it
gets kicked out of local minima after some time and continues its
path down (together with some kick-back mechanism\footnote{Trying to
minimize $V(\phi)^2$ instead will not work in this approach: $V^2$
has many codimension one loci of unwanted local minima, namely the
zero set of $V$. This can be avoided however in the discretized
version of the algorithm where we stick to the actual minima and
sequentially flip components of $m$.} when $V$ drops below zero, to
keep the energy positive), and indeed such procedures can be shown
to eventually do the job, but again only after exponentially long
time. Such algorithms are nevertheless often very useful in
practice, especially if one is satisfied with an approximate
solution. We will discuss them in more detail in the next section.

This example illustrates a deep conceptual connection between
\NP-hardness and physics: when a certain problem is \NP-hard,
\emph{any} ``physical'' way of solving it by rolling over some
effective potential landscape or by some other local relaxation
mechanism will come with exponentially many local minima, leading to
exponential relaxation times to the lowest lying minima.
This has immediate dynamical
implications for real physical systems: even when there is a
selection principle singling out a preferred state, such as minimal
energy, if the problem that needs to be solved to satisfy this
selection criterion is \NP-hard, the system may well be unable to
reach this preferred state on any reasonable time scale. This is
true for instance for spin glasses. We will look at this and its
consequences for fundamental physics in much more detail in
subsequent sections.

\subsection{Computational complexity of Bousso-Polchinski}
\label{subsec:compBP}

We now return to the complexity analysis of the Bousso-Polchinski
model. The problem we want to solve is the following: Does there
exist an $N \in \IZ^K$ such that
\begin{equation}
 \Lambda_1 \leq \Lambda_0 + \sum_{ij} g_{ij} N^i N^j \leq \Lambda_2?
\end{equation}
Here $K$, $g_{ij}$, and $\Lambda_r$ are all part of the input (and
$\Lambda_0 < 0$).

We prove that this problem is \NP-complete by relating it to a
version of the subset sum problem.\footnote{We thank Dieter Van
Melkebeek for suggesting this approach to us.} As in appendix
\ref{app:knapsackalg}, since we work at some finite precision, the
values of $\Lambda_1$, $\Lambda_2$ and $g_{ij}$ can all be assumed
to be integral after a suitable choice of energy units.

To show that this problem is \NP-complete, we have to show that some
known \NP-complete problem can be reduced to it in polynomial time.
The problem we choose for this is a version of subset sum with some
extra information (a ``promise'') about the input. The question we
ask is, given a set of $K$ positive integers $\{x_i\}$ and an
integer $t$, do there exist $k_i \in \{0,1\}$ such that $ \sum_i k_i
x_i = t$? The extra information is that the $x_i$ are such that
$\sum_i k_i x_i \neq t$ for any $k_i \in \IZ^+$, unless all $k_i \in
\{0,1\}$. In other words, in this problem we are promised that the
target $t$ cannot be reached by summing with multiplicities. The
\NP-completeness of this promise problem can be derived from the
\NP-completeness of the standard subset sum problem. The proof is a
bit technical, and we defer it to appendix \ref{app:NPCprom}.

Building on this, the reduction to the Bousso-Polchinksi problem is
straightforward: just take $g_{ij} = x_i \delta_{ij}$,
$\Lambda_1=\Lambda_2=t$. Thanks to the promise on the input, we know
that the fluxes $N_i$ can only take values $0$ or $\pm 1$ in any
solution of this problem, and therefore this instance of the BP
problem becomes identical to the above version of the subset sum
problem. This completes the proof.

The occurrence of exponentially many local minima in a local
minimization approach to the problem can be seen as follows. Assume
we take target values $\Lambda_1$ and $\Lambda_2$ to be very near 0,
say within $\epsilon \ll g$, where $g$ is the scale of $g_{ij}$. For
simplicity we take $g_{ij} = g_i \delta_{ij}$. We can try to reach
the target interval by starting at some large $N$ and minimizing
\begin{equation}
 |\Lambda|=|\sum_i g_{i} (N^i)^2 + \Lambda_0|.
\end{equation}
A single step in this procedure would consist of changing a random
component $N_k$ of $N$ by one unit: $\Delta N_i = \pm \delta_{ki}$.
This changes $\Lambda$ by an amount
 $$
  |\Delta \Lambda| = g_k |1 \pm 2 N^k| > g_k.
 $$
The minimization procedure will thus quickly relax down $|\Lambda|$,
but as soon as a value $|\Lambda| < g_k/2$ is reached, any
subsequent elementary step will in fact \emph{increase} $|\Lambda|$.
That is, we are at a local minimum. At this point, if $\epsilon \ll
\min_k g_k$, we are still very far from the target interval, and
moreover there are clearly exponentially many such local minima.
Again, we can try to further progress down by adding thermal noise,
but as we come closer and closer to $\Lambda = 0$, finding still
smaller values of $|\Lambda|$ (if they still exist) becomes
increasingly hard, as the smaller values will get further and
further away from each other. For the tiny values of $\epsilon$ we
are interested in, it will thus take exponentially many jumps to get
in the target range.

One could also consider the ``algorithm'' followed by cosmological
relaxation mechanisms proposed in this context
\cite{Brown:1987dd,Brown:1988kg,Bousso:2000xa,Feng:2000if}. We will
analyze these and discuss the corresponding implications of
computational complexity in detail in \cite{paperII}.

\subsection{Other lattice problems}

\label{subsec:otherlattice}

The Bousso-Polchinski problem is similar to well known
lattice problems such as the \emph{shortest lattice vector problem}
(SVP): given a lattice in $\IR^N$, find the shortest vector. It is
less obvious that this problem\footnote{or more precisely its
decision version; we will be a bit sloppy in our terminology in this
section.} is \NP-complete. For example, unlike in the BP problem, a
minimization algorithm would not need to explore increasingly
further apart lattice points. And indeed in the diagonal $g_{ij}$
case the problem is trivially in \P, the minimum length being
$\min_k g_k$. In fact, the \NP-hardness of this problem was long an
open question, but in 1998, Ajtai \cite{ajtaiSVP} proved it to be
\NP-hard under randomized reductions (which is slightly weaker than
standard \NP-hardness). Many approximation algorithms are known, most
famously the LLL algorithm \cite{LLL}, which finds a short vector in
polynomial time that is guaranteed to be at most a factor $2^{K/2}$
longer than the actual shortest vector. Various lower bounds on
polynomial time approximability are known, see e.g.\
\cite{Johnsoncolumn} for an overview.

Another well known \NP-hard lattice problem is the \emph{closest
lattice vector problem}: given a lattice and a point in $\IR^K$,
find the lattice point closest to that point. There are many other
hard lattice problems. In particular, a remarkable ``0-1 law''
conjecture by Ajtai \cite{ajtai01conj} produces a huge set of
\NP-complete lattice problems. The conjecture roughly says that any
polynomial time verifiable property of a lattice becomes generically
true or generically false for random lattices in the large lattice
dimension $K$ limit. In other words, only properties which in the
large $K$ limit can be statistically excluded or statistically
guaranteed can actually be possibly verified in polynomial time. Any
property that would be somewhat restrictive but not too restrictive
would automatically be intractable. An example is the question
whether there is a lattice point in some given region of volume not
much smaller or larger than the volume of a lattice cell. The
probability that there is such a point remains bounded away from 0
and 1 when $K \to \infty$, so if the conjecture holds, answering
this membership question is a problem not in \P. This automatically
includes SVP, CVP and BP. The conjecture actually also implies that
$\NP \neq \P$, so there is not much hope of proving it any time soon.

In the previous subsection we considered the problem of trying to
match the cosmological constant. One could consider different
parameters, such as particle masses, Yukawa couplings, and so on.
Experimental bounds on these parameters will typically map out some
finite size region in the space in which the flux lattice lives.
Hence Ajtai's conjecture implies that finding BP flux vacua
satisfying these constraints will in general be an \NP-hard problem.

\subsection{F-theory flux vacua and other combinatorial problems
 in string theory}

\label{subsec:Ftheory}

The Bousso-Polchinski model's main physical weakness is that it
ignores all moduli dependence of the potential.  From a computational
point of view, it does not explain the origin of the parameters 
$g_{ij}$ of the lattice, which are in fact not free parameters in
string theory.  One might wonder if the actual instances of BP
which arise are simpler than the worst case we discussed.

The next step in doing better is to construct
IIB superstring flux vacua, or more generally F-theory flux vacua
\cite{Dasgupta:1999ss,Giddings:2001yu}. Classically, one gets in
this setting a discretuum of supersymmetric vacua with all complex
structure (or shape) moduli stabilized, but the K\"ahler (or size)
moduli unaffected. As pointed out in \cite{Kachru:2003aw}, taking
into account quantum effects can supersymmetrically stabilize the
K\"ahler moduli as well, leading to vacua with negative cosmological
constant and no massless scalars. This was subsequently confirmed in
examples in \cite{Denef:2004dm,Denef:2005mm}, and extended to
nonsupersymmetric vacua with negative cosmological constant and
exponentially large compactification volume in
\cite{Balasubramanian:2004uy,Conlon:2004ds,Balasubramanian:2005zx}.
A plausible construction to uplift these vacua to positive
cosmological constant values was also proposed in
\cite{Kachru:2003aw}, and variants thereof in
\cite{Burgess:2003ic,Saltman:2004sn,Denef:2004cf}.

A good zeroth order approximation to the study of the landscape of
such flux vacua is to ignore the K\"ahler moduli altogether, and
consider the vacua of the potential on the complex structure moduli
space only, along the lines of
\cite{Ashok:2003gk,Denef:2004ze,Denef:2004cf}.

To sketch the actual problem that arises in this fully string
theoretic problem, and to show its relation to the idealized BP
model, let us introduce some formalism (this is not important for
subsequent sections however). F-theory flux is given by a harmonic
4-form $G$ on an elliptically fibered Calabi-Yau 4-fold $X$. The
flux $G$ is uniquely determined by its components with respect to an
integral basis of harmonic 4-forms $\Sigma_i$, $i = 1,\ldots,K$:
\begin{equation} \label{Gexp}
 G = N^i \Sigma_i,
\end{equation}
where $N^i \in \IZ$, because of Dirac quantization. We can also add
mobile D3-branes to the compactification. The four dimensional
effective potential\footnote{Consistent with our zeroth order
approximation, we neglect an overall factor depending on the
compactification volume (which is a K\"ahler modulus), and we
neglect warping effects.} induced by curvature, mobile D3-branes and
flux is, in suitable units:
\begin{equation} \label{eq:fluxpotdef}
 V = - \frac{\chi}{24} + N_{D3} + \frac{1}{2} \int_X G \wedge *G,
\end{equation}
where $\chi$ is the Euler characteristic of $X$ and $N_{D3}$ the
number of mobile D3-branes. Defining $\Lambda_0 \equiv -\chi/24 +
N_{D3}$ and $g_{i j} \equiv \frac{1}{2} \int_X \Sigma_i \wedge
* \Sigma_j$, and using (\ref{Gexp}), this becomes
\begin{equation}
 V = \Lambda_0 + g_{ij} N^i N^j .
\end{equation}
This is the same as the defining equation
\eq{BPmodeldef} of the BP model.  However, the main difference is that
the metric $g_{ij}$ depends on an additional set of complex variables 
$z^a$, with $a=1,\ldots,h^{3,1}(X)$.  Given a specific choice of $N^i$,
their values are determined by minimizing the energy \eq{BPmodeldef}.
This need only be a local minimum, so in general a choice of vacuum
is now a choice of $N^i$ and choice of minimum.

The source of this additional structure is that Ricci-flat metrics on
a Calabi-Yau manifold come in continuous families, in part
parameterized by the variables $z^a$ (the complex structure moduli).
Physically, such parameters lead to massless fields and long-range
forces, which are typically in conflict with the data, so this is a
problem.  However, in the presence of flux, the potential energy
\eq{fluxpotdef} depends on the $z^a$, as the choice of CY metric
enters into this expression through the Hodge star operator, so the
standard principle that energy is minimized in a vacuum fixes these
continuous variables and solves this
problem (this is the major reason for the physical interest in this
construction).  While the presence of parameters such as the $z^a$
is not obviously necessary to get a large vacuum
multiplicity, it is true of all known models which have it.

There are various further subtleties not taken into account in the BP model.
One is that the fluxes are constrained to satisfy the D3 charge
cancellation condition,
\begin{equation} \label{eq:tadpole}
 - \frac{\chi}{24} + N_{D3} + \frac{1}{2} \int G \wedge G = 0.
\end{equation}
Using this, one can rewrite \eq{fluxpotdef} as
\begin{equation}  \label{VGGdual}
 V = \frac{1}{4} \int_X (G-*G) \wedge *(G-*G).
\end{equation}
This is still not very explicit, due to the presence of the
$*$-operator, but it can be shown that this equals
\begin{equation} \label{eq:sugra}
 V_N(z) = e^{\CK} (G^{A{\bar B}} D_A W \bD_{\bar B} \bW - 3 |W|^2)
\end{equation}
where \cite{Gukov:1999ya}
\begin{equation}
 W_N(z) = N^i \Pi_i(z), \qquad \CK(z,\bz)=
 \Pi_{i}(z) Q^{ij} \bPi_{j}(\bz).
\end{equation}
Here $\Pi_i(z) = \int_{\Sigma} \Omega(z)$ is the holomorphic period
vector of the holomorphic 4-form, $G^{A\bar{B}}$ is the inverse
metric on moduli space, $D_A$ are compatible covariant derivatives,
and $Q_{ij} \equiv \int_X \Sigma_i \wedge \Sigma_j$ is the
intersection form on $H^4(X)$ and $Q^{ij}$ its inverse.
We omit further details, which can be found in
\cite{Giddings:2001yu,Kachru:2003aw,%
Douglas:2005df,Denef:2004ze,Denef:2004cf}
and the other references.\footnote{
For those more familiar with this problem, the indices
$A,B$ range over both complex and K\"ahler moduli; the
reintroduction of K\"ahler moduli at this level has as only effect
to cancel off the negative contribution to $V_N$, in accord with the
positive definite (\ref{VGGdual}). However, after including quantum
corrections \cite{Kachru:2003aw}, equation (\ref{VGGdual}) no longer
holds, while on general grounds (\ref{eq:sugra}) is still valid, but
now the negative term is no longer cancelled off identically, and
therefore the potential will not be positive definite. In fully
stabilized models taking into account quantum corrections one can
therefore expect (\ref{eq:sugra}) with $A,B$ ranging over just the
complex structure moduli to be a better model for the complex
structure sector than the same omitting the negative term, as it
would in the very special classical case.}

The main point is that this part of the problem is mathematically
precise and sufficiently concrete to make explicit computation of the
potential $V_N(z)$ possible in examples for all choices of $N$ and
$z$.  There are various approaches to computing the data we just
described; for example the periods are determined by a Picard-Fuchs
system of partial differential equations, which given a choice of
Calabi-Yau manifold contains no adjustable parameters.  In concrete
constructions, the moduli at a minimum of $V$ control observables such
as coupling constants and masses of particles, and this model is
perhaps the most fully realized example to date of how we believe
string theory can in principle determine all the continuous parameters
of the Standard Model from a starting point with no parameters, again
always under the assumption that we know which vacuum to consider.

Given this setup, one can again ask for the existence of vacua with
cosmological constant in a given small range. This problem is similar
in spirit to the Bousso-Polchinski problem, but appears harder because
of the coupling to the moduli, which will have different critical
point values for each choice of flux. Therefore one would expect this
problem to be at least as intractable, and no algorithm to exist that
would guarantee a solution of the problem in polynomial time.

If we restrict attention to actual Calabi-Yau compactifications, then
strictly speaking it does not really make sense to call this
problem \NP-hard, since \NP-hardness is an asymptotic notion, and there
are reasons to think that only a finite number of instances can actually
be produced in string theory \cite{Douglas:2005hq}.
However, since the number of fluxes on
elliptically fibered Calabi-Yau fourfolds can be at least as large
as about 30,000 (and as large as about 2,000,000 for general Fermat
fourfold hypersurfaces) \cite{Klemm:1996ts}, there is clearly little
reason to doubt the effective intractability of the problem, at
least in its current formulation.

Let us suggest a version of this problem with an asymptotic limit, in
which the complexity question is well posed.  As discussed for example
in \cite{Douglas:2005df}, one does not need an actual Calabi-Yau
manifold $X$ to pose this problem, merely a ``prepotential,'' a
holomorphic function of $K/2-1$ complex variables $z^i$, which
summarizes the geometric information which enters the problem.  Thus,
rather than input a lattice as in BP, we input a prepotential and a
number $\chi$ as in \eq{tadpole}, and ask whether the resulting set of
flux vacua contains one with $V$ as defined in \eq{sugra} in a
specified range.  Since we are not asking for the prepotential to
correspond to an actual Calabi-Yau manifold, the problem size $K$ can
be arbitrarily large.  One might also propose variations on this
which capture more structure of the actual problem, such as to base
the construction on
variation of Hodge structure for some infinite set of manifolds.

Explicit algorithms to solve any of these problems, even
approximately, would be of real value to string theorists.  However,
there is ample scope for reduction arguments which might prove their
\NP-completeness as well.  For example, the Taylor series expansion of
the prepotential contains far more information than a metric $g_{ij}$
specifying a $K$-dimensional lattice, suggesting that a similar
(though far more intricate) reduction argument could be made.

As discussed in the previous subsection, similar consideration hold
for matching other continuous parameters, such as particle masses
and Yukawa couplings. What about discrete quantities, such as gauge
groups, numbers of generations and so on? Here again one encounters
\NP-hard problems: typically one needs to find all D-brane
configurations in a given compactification consistent with tadpole
cancellation and the discrete target requirements, which is very
similar to subset sum problems. Such a combinatorial D-brane problem
was studied in detail in a simple model in
\cite{Blumenhagen:2004xx,Gmeiner:2005vz}.
In \cite{Gmeiner:2005vz}, it was suggested 
(without giving precise arguments) 
that this problem (in its asymptotic
extension) is indeed \NP-hard. Because of this, the authors had to
resort to an exhaustive computer search of a large subset of
solutions, scanning about $10^8$ models, which required $4 \times
10^5$ CPU hours. To increase a certain measure of the problem size
by a factor of 2, they estimated they would require a computation
time of $10^8$ CPU \emph{years}. This makes the practical hardness
of the problem quite clear.

Obviously, this suggestion would not mean that one cannot construct
particular instances of models with, say, three generations.  One
might also hope to solve important pieces of the problem in polynomial
time.  But it would imply that one cannot construct general algorithms
that systematically examine all solutions, finishing in guaranteed
polynomial time (unless, of course, \P\ turns out to equal \NP).

We should also emphasize that \NP-hardness does not mean that any
instance of the problem will take exponential time to solve.
\NP-hardness is strictly speaking a worst case notion. Many
instances may be easy to solve; for example we saw that the knapsack
problem with a target range that is not too small compared to the
typical size of the entries is effectively solvable in polynomial
time.\footnote{
Actually, the well studied lattice problems such as SVP and CVP are
not just worst case hard, but hard for average instances, a fact
exploited in cryptography.  It would be interesting to check
whether BP shares this property.}
In general, when there is an exponentially large number of
vacua satisfying the target constraints, finding one of them will be
relatively easy. But such cases are of limited interest: physically,
we want the constraints to be sufficiently tight to select only one
or a few vacua, or at most select a manageable and enumerable set,
and finding those tends to be exponentially hard for \NP-hard
problems. So we get a complementarity: predictivity versus
computational complexity. The more selective a criterion, the harder
it will be solve the associated selection problem.

However, when the selection criteria get so restrictive that one
does not expect any solutions at all within a given ensemble (e.g.\
on statistical grounds), the problem of answering the question
whether there are indeed no solutions may sometimes get much easier
again. A trivial example is the subset sum problem for a target
value close to the sum of all positive integers in the given list --
one needs to check at most a few cases to solve this case. Thus,
\emph{excluding} certain ensembles of models may on general grounds
still be a tractable task.

Finally, we note that there are often other, more efficient ways to
extract physical predictions besides explicitly solving selection
problems. There are plenty of complex physical systems for which
finding the microscopic ground state or some set of microscopic
states satisfying a number of macroscopic criteria is completely
intractable, and yet one can learn a lot about the physics of these
systems using statistical mechanics, even with only rough knowledge
of the microscopic dynamics. Particularly relevant here is the
theory of spin glasses, to which we return in section
\ref{sec:relprobs}. Statistical methods to analyze the landscape of
flux vacua were developed in
\cite{Ashok:2003gk,Denef:2004ze,Denef:2004cf}, and we will discuss
how to address this problem in that context in section 
\ref{sec:practical}.

\subsection{Beyond toy models}

Our discussion so far concerns toy models of the real string theory
landscape, which are relatively rough approximations to the exact
string theoretic problems. Even granting that these properly reflect
the situation, or at least give a lower bound on the actual
complexity of the landscape, we should discuss the added
complications of more realistic problems. One obvious issue is the
precision at which we can actually compute the cosmological constant
and other parameters in a given string vacuum. If this is
insufficient, we are in practice simply not in the position to even
try to solve problems like those presented above.

We first note that there are string theoretic problems of the same
general type we are discussing, in which it is known how to compute
exact results, so that the discussion is precise.  For example, we
have the problem of finding a BPS black hole in IIb string theory on
a Calabi-Yau threefold, whose entropy satisfies specified bounds.
The entropy is determined by the charges $N^i$ through the attractor
mechanism \cite{Ferrara:1995ih}; it is the minimum of a function of
the form $S=e^{K}|N^i \Pi_i|^2$. This is problem is very similar to
the F-theory flux vacua problem outlined above. However in the black
hole case there are no approximations involved; everything is exact.

At present all such exactly solvable problems assume supersymmetry,
and at least eight supercharges, so that one can get exact results
both for the superpotential and K\"ahler potential.  While one can
hope for a similar level of control in $\CN=1$ supersymmetric vacua
in the foreseeable future, it will be a long time
before we have the computational ability to compute the cosmological
constant to anything like $10^{-120}$ accuracy in even a single
non-trivial example with broken supersymmetry.  This will become
clear after we outline how this is done in section \ref{sec:practical}.

If we were to grant that this will remain the permanent situation,
then by definition the problem we are posing is intractable; no
further arguments are required.  However, there is no principle we
know of that implies that this must remain so.  Well controlled series
expansions or even exact solutions for many physical problems have
been found, and although looking for one here may seem exceedingly
optimistic from where we stand now, who can say what the theoretical
situation will be in the fullness of time. Our point is rather that,
granting that the large number of vacua we are talking about actually
exist in the theory, presumably the natural outcome of such
computations will be a list of numbers, the vacuum energies and other
parameters in a large set of vacua, and that some sort of search of
the type we are discussing would remain. Barring the discovery of
extraordinary structure in the resulting list, the present evidence
suggests that at this point one would run into a wall of computational
complexity with its origins in deep theorems and conjectures of
theoretical computer science, rather than just technical limitations.

\section{Related computational problems in physics}
\label{sec:relprobs}

There is a large literature exploring the relations between
physics and computation, let us mention
\cite{Bennett,Feynman,Deutsch,Preskill}.
What is of most relevance for us here is the particular
case of finding the ground state(s) of complex physical systems.
This has an even larger literature, both because of its physical and
technological importance, and because it provides a particularly natural
encoding of computational problems into physics.

\subsection{Spin glasses}

A prototypical example is the Sherrington-Kirkpatrick (SK) model of a spin
glass \cite{SK}.
A spin glass is a substance which contains dilute magnetic spins
scattered about the sample, so that the individual spin-spin couplings
vary in a quasi-random way.  This can be modeled by
a statistical mechanical system whose degrees of
freedom are $N$ two-valued spins $\sigma_i=\pm 1$, and the Hamiltonian
\begin{equation} \label{eq:spinglassH}
H = \sum_{1\le i<j\le N} J_{ij} \sigma_i \sigma_j .
\end{equation}
In this model, the statistical sum over states is
just the sum over spins; the partition function for a single sample is
\begin{equation} \label{eq:spinZ}
Z = \sum_{\{\sigma_i=\pm 1\}} e^{-\beta H}
\end{equation}
at fixed $J_{ij}$.  However, the couplings $J_{ij}$ are not known
{\it a priori}, as they vary from sample to sample.

Rather, we study an ensemble
of spin glasses, specified by giving a probability distribution
$[DJ]$ on the couplings.  An expectation value in such an ensemble is
defined as
$$
\langle\langle X \rangle\rangle \equiv
\int [DJ]\  \frac{1}{Z}
 \sum_{\{\sigma_i=\pm 1\}} e^{-\beta H} \ X .
$$
where $Z$ is as in \eq{spinZ}.  Although this expression averages over
both the spins $\sigma$ and the couplings $J$, they are treated
differently -- while the $\sigma$ weights depend on $J$,
the average over $J$ is done with the dependence on $\sigma$ removed.
This is because these couplings were
fixed during the preparation of the sample,
long before any spin dynamics came into play.
This is known as quenched disorder.

The SK model is defined by taking the couplings $J_{ij}$ to be
independent random variables, drawn from a Gaussian distribution with
variance $1/N$.  This normalization is made so that the free energy is
$O(N)$ in the thermodynamic limit.
Compared to a real spin glass, the main simplification
of the SK model is that we have neglected the spatial distribution,
in effect taking the limit of spins in infinite spatial dimensions.

Besides numerical study, some analytic information about
such systems can be found by various tricks, the most general
of which is the ``replica trick'' as discussed in \cite{Mezard}.
The picture of the SK model
which emerges is that, as a function of inverse temperature
$\beta$, there are two regimes.  At high temperature, the interactions
are unimportant, and the spins are disordered.  Conversely, at low
temperature, there is ordering, but not detectable by a simple order
parameter such as the average expectation value of the spins.
It can be detected, for example, by the
Edwards-Anderson order parameter
$$
q_{EA} \equiv \frac{1}{N}\sum_{i=1}^N \vev{\sigma_i}^2
$$
Furthermore, unlike an Ising model (the case with constant $J_{ij}<0$),
the partition function is not dominated by the
lowest energy spin configuration, the global minimum of $H$.
This is because of the existence
of a large number of local minima (or ``vacua'').  These arise because
of frustration.

Frustration refers to the fact that, for generic couplings $J_{ij}$, one will
find many cases of $1\le i,j,k\le N$ for which
$$
J_{ij} J_{jk} J_{ki} < 0
$$
and thus there is no preferred assignment of the spins $\sigma_i$,
$\sigma_j$ and $\sigma_k$ which will minimize all three of their
pairwise interaction energies.  In this situation, stepwise lowering
the energy of spin configurations tends to end in local minima, in
which flipping one or a few spins raises the energy, but by flipping
many spins one can reach a new potential well with a very different
set of competing interactions and a different value of the potential
at the minimum.

It is intuitively clear that the phenomenon of frustration makes the
problem of finding the absolute minimal energy ground state for
given couplings $J_{ij}$ computationally hard, and indeed it has
been shown that the problem is \NP-hard \cite{Barahona}.

Note the general similarity to the Bousso-Polchinski model. In
particular, both models share the basic feature of possessing an
exponentially large number of ``vacua,'' in BP because varying a
single flux produces a large variation in energy, and in SK because
of frustration.

As we mentioned earlier, this means that physical models typically
will not find the global minimum within the relevant time scales,
and this has real physical consequences for spin glasses.  A spin
glass magnetized at a high temperature in the presence of an
external magnetic field shows a very characteristic relaxation
behavior when cooled below its critical temperature and taken away
from the external field. After a quick initial drop of magnetization
to some characteristic value, it starts an extremely slow descent
towards zero (or at least very small) magnetization. In fact it
never quite reaches thermal equilibrium within experimentally
accessible time scales. The reason is that the system keeps on
getting stuck in the exponentially large number of local metastable
minima of its energy landscape, much like a computational relaxation
algorithm would. As a result, the properties of such systems depend
significantly on their preparation history and the waiting time:
they are ``aging.''
This implies something rather remarkable, not shared by
computationally simple systems such as ferromagnets or regular
crystals, namely that just from observation of the system's current
state, we can infer some knowledge of its history.

\subsection{Physically inspired methods for finding the ground state}

Let us consider some ways in which one might try to find the ground
state. The simplest is to choose a starting configuration and evolve
by flipping single spins, accepting any which decrease the energy.
This will rapidly stop at a local minimum.  Physically, it
corresponds to rapidly cooling the sample to zero temperature.

A more sophisticated approach which can deal with local minima is
simulated annealing \cite{Sim}.  In this approach, we simulate the system
according to the standard Metropolis algorithm for finite
temperature. We recall that the Metropolis algorithm also proceeds by single
spin flips, accepting any flip which lowers the energy, and
accepting flips which raise the energy by $\Delta H$ with
probability $\exp(-\beta\Delta H)$, thus generating the Boltzmann
distribution.

Simulating annealing is then a Metropolis process which starts at high
temperature, able to explore the entire configuration space, and then
systematically lowers the temperature to zero.  This process allows jumping
out of local minima, while by taking the temperature
to zero sufficiently slowly, one is confined to potential basins of lower
and lower energy, eventually finding the global
minimum.  However, it has been argued \cite{Geman} that to guarantee this
requires taking $\beta\sim \log t$ at time $t$, {\it i.e.}
the total process takes exponential time, just as would all known
algorithms on a classical computer.

Again, these are worst case results, while in practice some problems
can be solved.  In these cases, clever algorithms can provide
significant speedups.  For example, spin glass techniques inspired the
``cluster method'' of M\'ezard et al \cite{Mezard}, and many other
methods such as deformation, basin hopping, and $\alpha$ branch and
bound, have been found useful in practical optimization problems
\cite{Wales}.

\subsection{Reformulation of other \NP-complete problems as physical systems}

As mentioned earlier, it has been shown that the problem of finding
the ground state of the spin glass \eq{spinglassH} with specified
couplings $J_{ij}$ is \NP-hard \cite{Barahona}. This was done by
relating this to a graph theoretic problem which was known to be
\NP-hard.

A more natural result of this type is the mapping of the
satisfiability problem into a spin glass with multi-spin interactions.
This is very simple and is the starting point for a lot of work using
statistical mechanical approaches to hard optimization problems.

Recall that in the $3$-\SAT\ problem, one is given a list of
Boolean OR clauses, for example
$$
x_2 \vee \bar{x_7} \vee x_{22}
$$
where the bar represents NOT,
and $\vee$ represents OR.  The problem is to find an assignment of
truth values to variables which satisfies all of the clauses.

This is trivially identical to the problem of
finding a zero energy state of the following potential,
$$
V =  \sum_{i=1}^N
 \left(\frac{1+ J_{ia} \sigma_a}{2}\right)
 \left(\frac{1+ J_{ib} \sigma_b}{2}\right)
 \left(\frac{1+ J_{ic} \sigma_c}{2}\right)
$$
where $\sigma_a=\pm 1$ represent the variables, and
the coefficients $J_{ia}, J_{ib}, J_{ic}=\mp 1$ encode the choice of
a variable or its negation in the $i$'th clause.  If any of
$J_{ia}\sigma_a=-1$, the clause is satisfied and has zero energy, while if
all
$J_{ia}\sigma_a=+1$, the clause is not satisfied and has positive energy.

Various interesting insights were obtained by using this mapping as
well as the general relations between optimization and statistical
mechanics.  As an example which inspired much interest \cite{Sat}, if
we consider a randomly chosen $3$-SAT problem (with randomly chosen
$J_{ia}$) and vary the ratio $\alpha$ between the number of clauses
and the number of variables, there appears to be a phase transition: for
$\alpha<\alpha_c\sim 4.267$ the problems are satisfiable (meaning that
as $N\rightarrow \infty$ the fraction of satisfiable problems goes to
$1$), while for $\alpha>\alpha_c$ they are not.

This phase transition is directly related to the difficulty of the
problem.  Finding an explicit solution is easy for small $\alpha$, and
becomes more difficult as constraints are added.  Conversely, the
difficulty of showing that no solution exists (a very different
problem, as we discuss later) decreases with $\alpha$.  In some sense,
the overall difficulty of the problem peaks at $\alpha=\alpha_c$.

\subsection{Protein landscapes} \label{ss:lands}

Another much studied example is the potential landscape provided by
configurations of proteins. The problem of finding the folded ground
state of a protein (modeled by various discretized models) is known
to be \NP-hard \cite{unger}, and simulations of protein folding
based on these models suffer from the usual problem of getting stuck
in metastable energy minima, making the problem computationally
intractable already for relatively short sequences of amino acids.
Again, the hardness of the problem has physical implications.
Artificially made random sequences of amino acids generically do not
fold properly: they do not find a unique folded ground state, as one
would expect based on the \NP-hardness of the problem. However, the
story is quite different for biologically occurring proteins, which
typically fold into a unique preferred ground state very quickly
after being produced in a cell. These native states tend to be very
stable, and proteins that are denaturated (that is, unfolded) by
heating or chemical reactions often have no trouble folding back
into their native state. Given the apparent computational complexity
of the problem, this presents a puzzle, referred to as
\emph{Levinthal's paradox} \cite{Levinthal}.

The resolution of this paradox is evolution: the processes involved
in synthesizing proteins, and in particular the actual amino acid
sequence itself \cite{Socolich} have been selected over billions of
years and a huge number of trials to be exactly such that biological
folding is efficient and reliable. The particular landscape and
folding pathways of natural proteins are such that it is effectively
funneled into a unique native state. Failure to do so would result
in dysfunctional protein and weakening or elimination of the
organism. In other words, whereas computational complexity is a
notion based on worst case instances, there is strong evolutionary
pressure to select for \emph{best} case instances.\footnote{But
which instances these are depends on the details of the dynamics (in
other words the algorithm), and \emph{finding} these best case
instances is conceivably again computationally hard. However, the
mechanism of evolution provides enormous space and time resources to
do this.}

In a way, the intractability of the general problem is
again what allows the system to carry information about the past, in
this case the whole process of evolution.

\subsection{Quantum computation} \label{subsec:quantum}

It is interesting to ask whether using a quantum computer brings any
speed-up in solving these problems, and this will be especially
interesting for the second paper.
For background on quantum computing, consult
\cite{Kitaev,Nielsen}.

Although there are many approaches to performing a computation using
quantum mechanics, the most relevant for our discussion here is to
translate it into the problem of finding a ground state of a quantum
mechanical system, along the general lines we just discussed.  The idea of
doing computations this way gains particular interest from the concept
of adiabatic quantum computing \cite{Farhi}.  The idea is to find the
ground state by starting from the known ground state of a simple
Hamiltonian, and varying the Hamiltonian with time to one of interest.
According to the adiabatic theorem of quantum mechanics, if this is
done slowly enough, the original ground state will smoothly evolve
into the ground state of the final Hamiltonian, and thus the solution
of the problem.  In fact any quantum computation can be translated
into this framework, as recently shown by Aharonov {\it et al}
\cite{Aharonov}.

Similar to the classical case, one defines complexity classes for
problems solvable by quantum computers with specified resources. Of
course for a quantum computer, all computation outputs are
probabilistic, so all classes will have to specify the error
tolerance as well. The class of problems viewed as tractable by a
quantum computer is \BQP, or Bounded-error Quantum Polynomial time.
This is the class of decision problems solvable by a quantum
computer in polynomial time, with at most 1/3 probability of error.
The number 1/3 is conventional, any nonzero $\epsilon < 1/2$ would
give an equivalent definition. If the probability of error is
bounded in this way, one can always reduce the probability of error
to an arbitrarily small value by repeating the computation a number
of times, where the required number depends only on the desired
accuracy.
%The analogous classical complexity class is \BPP\ (Bounded error
%Probabilistic Polynomial time), problems solvable in polynomial time
%by a classical randomized algorithm with at most 1/3 probability of
%error.
Many results relating classical and quantum complexity classes are
known, see \cite{complexityzoo}.

To admittedly oversimplify a complex and evolving story, while there
are famous examples of problems for which quantum computers provide an
exponential speedup, such as factoring integers \cite{Shor}, at present
the evidence favors a simple hypothesis according to which a generic
problem which takes time $T$ for a classical computer, can be solved
in time $\sqrt{T}$ by a quantum computer.\footnote{
As pointed out to us by Scott Aaronson, this hypothesis is simplistic,
as there are also problems with no asymptotic speedup over classical 
computers, or with $T^{\alpha}$ speedup with $1/2<\alpha<1$.  
But perhaps it is a reasonable 
first guess for the search problems which concern us here.}
The simplest example of
this is the Grover search algorithm \cite{Grover}, and this result can
be interpreted as providing general evidence for the hypothesis by the
device of formulating a general computation as an oracle problem
\cite{Bennett}.

Many other cases of this type of speedup are known.  Another
relevant example is the problem of estimating an integral by
probabilistic methods.  As is well known, for a generic function with
$O(1)$ derivatives, the standard Monte Carlo approach provides an
estimate of the integral with $O(T^{-1/2})$ accuracy after sampling
$T$ points.  If we assume a function evaluation takes unit time, this
takes time $T$.  On the other hand, a quantum computer can use $T$
function evaluations to estimate the same integral to an accuracy
$T^{-1}$. \cite{Grover2}

While significant, against exponential time complexity, a square root
improvement does not help very much; an \NP-hard problem will still
take exponential time to solve.  This also seems to come out of the
adiabatic quantum computation framework, in which one constructs a
family of Hamiltonians which adiabatically evolves to a Hamiltonian
whose ground state solves an \NP-hard problem.  In the known examples,
such a family of Hamiltonians will contain excited states with
exponentially small gap above the ground state, so that the time
required for adiabatic evolution is exponentially long  (see
\cite{Farhi2} and references there).

The problems for which quantum computation is presently known to
offer a more significant speedup are very special \cite{Shor2}.  Many
can be reformulated in terms of the ``hidden subgroup problem,'' which
includes as a special case the problem of detecting periodicity in the
sequence of successive powers of a number, exploited in Shor's
factoring algorithm.  Of course lattice problems have an underlying
abelian group structure as well and it is conceivable that quantum
computers will turn out to have more power here.\footnote{ A
primary application of lattice algorithms is to cryptography, and
we have been told that because of this, much of
this research literature is government classified.  For all we know,
the technology we need to find string vacua may already exist at the
NSA.}

To conclude, it would be very interesting to have precise statements
on the computational power of quantum field theory, compared to
generic quantum mechanical systems.  A precise discussion of this
point would also enable us to discuss interesting questions such as
whether computational power is invariant under duality equivalances
\cite{Preskill2}.  It has been studied in depth for topological
quantum field theory 
\cite{Freedman:2000rc}, but this is a rather special
case, since for any given observable one can reduce such a theory to
finitely many degrees of freedom.  In contrast, formulating a general
quantum field theory requires postulating an infinite number of
degrees of freedom, the modes of the field at arbitrarily large
energies.  On the other hand, one expects that for any observable,
there is some finite energy $E$, such that modes of larger energy
decouple, and only finitely many modes enter in a non-trivial way.
The question is to make this precise and estimate the number of
quantum computational operations which are available as a function of
physical resources time $T$, volume $V$ and energy $E$.

Locality and dimensional analysis suggest that a general upper bound
for the number of computations $N$ which can be done by a $d+1$
dimensional theory in time $T$ and in a region of volume $V$ should
take the form $N\le TVE^{d+1}$, where $E$ has units of energy.
However, it is not clear what determines $E$.  The masses of the
heaviest stable particles, other natural scales in the theory, and
properties of the initial conditions, might all play a role, and 
enter differently for different theories.

A closely related question is the difficulty of simulating a QFT by a
quantum mechanical computer; {\it e.g.} what is the number of quantum
gate operations required to compute the partition function or some
other observable to a desired accuracy.  The only directly relevant
work we know of is \cite{Byrnes}, which suggests that simulating a
lattice gauge theory with lattice spacing $a$ requires $TV/a^{d+1}$
computations, as one might expect.  However, as defining a continuum
QFT requires taking the limit $a\rightarrow 0$, this estimate is at
best an intermediate step towards such a bound.  One would need to use
the renormalization group or similar physics to summarize all the
dynamics at high energies by some finite computation, to complete this
type of analysis.

\section{Problems harder than \NP} \label{sec:harderprob}

We now return to our string theoretic problems. One response to the
difficulties of finding vacua with parameters such as $\Lambda$ in a
prescribed (e.g.\ by experiment) target range, as described in
section \ref{sec:fluxNP}, is to suggest that we have not taken into
account all of the physics of early cosmology, and that properties
of the initial conditions, dynamics or other effects will favor some
vacuum over all others. Perhaps the problem of finding this
``pre-selected'' vacuum will turn out to be much easier than the
problems described in section \ref{sec:fluxNP}. All we would have to
do then is to compute this preferred vacuum and compare to
observations.

Here we consider a simple candidate principle which actually does
this -- in principle.  As we will see in this section, 
trying to use it in practice leads to a computational problem
more intractable than \NP.  We continue with a survey of additional
concepts in complexity theory which will be useful in the sequel.

\subsection{Sharp selection principles based on extremalization}

\label{subsec:sharpselection}

What might be a principle which prefers or selects out a subset of
vacua?
From our present understanding of string theory, it seems unreasonable
to hope for a principle which {\it a priori} selects out (say) a
certain ten dimensional string theory, a particular Calabi-Yau
threefold, bundle and fluxes, and so on.  What seems more plausible is
a principle that gives us an amplitude or probability distribution on
the set of candidate vacua, which might be a function of their
physical parameters, the size of the set of initial conditions which
can evolve to the vacuum of interest, and so forth.  This is usually
referred to as a ``measure factor'' in the cosmological literature,
and the probability distribution of vacua with these weights is the
``prior distribution'' or simply the ``prior'' (as in Bayesian
analysis).

While one can imagine many possibilities, for definiteness let us
consider, say, the idea that only vacua with positive cosmological
constant can appear, and these with a probability which depends only
on the c.c., as
\begin{equation} \label{eq:ExpLam}
P(\Lambda) \propto e^{24 \pi^2 M_{P}^4/\Lambda} ;\qquad \Lambda>0
\end{equation}
for positive $\Lambda$, and probability zero for $\Lambda\le 0$. We
grant that the sum of these factors over all metastable vacua is
finite, so that this can be normalized to a probability
distribution. The exponent is the entropy $S(\Lambda)$ of
four-dimensional de Sitter space with cosmological constant
$\Lambda$  \cite{Gibbons:1976ue}.

This proposal has a long and checkered history which we will not try
to recount (we will give more details in \cite{paperII}).
As, taken at face value, it appears to
offer a solution to the cosmological constant problem, there are
many works which have argued for and against it, perhaps the most
famous being \cite{Hawking}. A simple argument for the proposal,
using only general properties of quantum gravity, is that it follows
if we grant that the number of microstates of a vacuum with
cosmological constant $\Lambda$ is proportional to $\exp S$, with
$S$ the dS entropy, and that these states undergo a dynamics
involving transitions between vacua which satisfies the principle of
detailed balance, as then this would be the expected probability of
finding a statistical mechanical system in such a macrostate.
%\cite{Banks:2005ru} + others probably such as Linde.

The proposal can be criticized on many grounds: the restriction to
$\Lambda>0$ is put in by hand, the relevant vacua in string theory
are probably not eternal de Sitter vacua, and so on (see however
\cite{Banks:2005ru} for recent, more detailed arguments in favor of
this proposal). Furthermore, some specific frameworks leading to the
proposal make other incorrect predictions. For example, the argument
we mentioned might suggest that the resulting universe would be in a
generic state of high entropy, predicting a cold and empty de Sitter
universe.

In any case, if we simply take the proposal at face value, it at
least makes a definite prediction which is not immediately falsified
by the existing data, and thus it seems a good illustration of the
general problem of using measure factors.

The measure factor \eq{ExpLam} is extremely sharply peaked near
zero, and thus for many distributions of $\Lambda$ among physical
vacua it is a good approximation to treat it as unity on the vacuum
with the minimal positive cosmological constant $\Lambda_{min}$, and
zero on the others.  To illustrate this, let us grant that the
distribution of cosmological constants near zero is roughly uniform,
as is reasonable on general grounds \cite{Weinberg:1987dv}, and as
confirmed by detailed study \cite{Bousso:2000xa,Denef:2004cf}. In
this case, one expects the next-to-minimal value to be roughly
$2\Lambda_{min}$, and the probability of obtaining this vacuum
compared to that of the minimal vacuum is of order
$\exp(-1/2\Lambda_{min})$, thus negligible. We will refer to
the special case of a measure factor which is overwhelmingly peaked
in this way as ``pre-selection.''

We point out in passing that, to the extent that we believe that the
cosmological evidence points to a specific non-zero cosmological
constant of order $10^{-120}M_{P}^4$, there is a simple independent
theoretical test of the proposal. It is that, now granting that the
distribution of cosmological constants is roughly uniform over the
entire range $(0,M_{P}^4)$, the total number of consistent
metastable vacua should be approximately $10^{120}$, since if it is
much larger, we would expect the cosmological constant of the
selected vacuum to be much smaller than the measured value (which
would be a rather ironic outcome, given the history of this
proposal). Granting this, the combination of these ideas would
provide a simple explanation for the observed value of $\Lambda$,
and in principle determine an overwhelmingly preferred unique
candidate vacuum.

However, before we begin celebrating, let us now consider the
problem of actually finding this preferred candidate vacuum.  Given
a concrete model such as BP, it is mathematically well posed; we
simply need to find the minimum positive value attained by the c.c..
However, as one might imagine, proving that one has found a minimal
value is more difficult than simply finding a value which lies
within a given range.  Whereas the latter condition can be verified
in polynomial time, here even verifying the condition would appear
to require a search through all candidate vacua.  Thus, apparently
this problem is not even in \NP.

To be more precise, we consider the decision problem \MINCC, defined
as the answer to the question,
\begin{quotation}
Does the minimal positive c.c. of the theory lie in the range
$[\Lambda_0-\epsilon,\Lambda_0+\epsilon]$?
\end{quotation}
Here $\Lambda_0$ could be the presently measured cosmological
constant and $\epsilon$ the measurement error. To stay within the
standard framework of complexity theory, we have formulated the
problem as a decision problem, rather than as the problem of
actually computing the minimal value. Note however that if we have
an oracle that answers this question in one step, we can bracket the
minimal value to a precision of order $1/2^n$ after $n$ steps. The
\MINCC\ problem is equivalent to a positive answer to both of the
following two decision problems:
\begin{enumerate}
 \item \CC: Does there exist a vacuum $v$ with $0 < \Lambda(v)
 \leq \Lambda_0 + \epsilon$ ?
 \item \MIN: For all vacua $w$, does either
 $\Lambda(w) \leq 0$ or $\Lambda(w) > \Lambda_0 - \epsilon$ ?
\end{enumerate}
While the first problem (\CC) is in \NP\ (assuming $\Lambda(v)$ can
be computed in polynomial time, as is the case in the BP model), the
second problem (\MIN) appears of a different kind, since a positive
answer to the question cannot be checked by a simple evaluation of
$\Lambda(v)$ for some suitable $v$. In fact, it is by definition a
problem in \coNP, the complementary class to \NP: the problem ``is X
true?'' is in \coNP\ iff the complementary problem ``is X false?''
is in \NP.\footnote{Recall there is an asymmetry between yes and no
in the definition of \NP: we only require a yes answer to be
verifiable in polynomial time.} In this case, the complementary
problem is
$$
 \exists w: 0 < \Lambda(w) \leq \Lambda_0 - \epsilon ~?
$$
which is clearly in \NP. More generally, we have that \NP\ problems
have the logical structure $\exists w: R(\epsilon,w)$, while \coNP\
problems have the structure $\forall w: R(\epsilon,w)$, where $R$ is
some polynomial time predicate.

Thus, the problem \MINCC\ is the conjunction of a problem in \NP\
and a problem in \coNP.  This is by definition in the class \DP\
(for ``Difference \P'').  An example of a universal (or complete)
problem in this class is to decide if the \emph{shortest} solution
to the traveling salesman problem is of a given length $l$.  While
one can clearly solve this problem in finite time (by enumerating
all solutions of length at most $l$), since it is not obviously in
\NP, it may be more difficult than problems in \NP.

Complexity theorists strongly believe that the class \DP\ is
strictly larger (and thus intrinsically more difficult) than either
\NP\ or \coNP. As with $\P\ne\NP$, this belief is founded on
experience with a large set of problems, and the consistency of a
world-view formed from results which bear indirectly on the
question.  Of course, the underlying intuition, that finding an
optimal solution should be harder than just finding a solution, is
plausible and this might be enough for some readers.  In the rest of
this subsection we go on and briefly describe one of the main
arguments for this, as explained in \cite{Rudich}.

A standard way to think about such problems in complexity theory is
to grant that we have an oracle which can solve some part of our
problem in a single time step.  For example, we might grant an
oracle which, given a candidate $v$, answers the question \MIN\ in
one step. Given such an oracle, the problem \MINCC\ is in \NP, as
the remaining problem \CC\ is in \NP. Such a ``relativized'' class
is denoted by superscripting with the class of the oracle, so the
problem \MINCC\ is in the class $\NP^{\coNP}$. This is much larger
than \DP, so at this point we have not learned much, but let us
continue.

Now, an \NP\ oracle can solve \coNP\ problems, and vice versa. To
see this, simply recall that by definition, the yes/no problem ``is
X true?'' is in \coNP\ iff the problem ``is X false?'' is in \NP. A
yes/no answer to the second question is also an answer to the first
question, so \NP\ and \coNP\ oracles are the same.

Thus, $\NP^{\coNP}$ is the same as $\NP^\NP$, which is also called
$\Sig2$. This class answers questions of the form
 $$
  \exists w_1 \forall w_2~ R(\epsilon,w_1,w_2).
 $$
A physics example of this would be: ``Is the height of the potential
barrier between two given local minima $x_i$ and $x_f$ at least
$\epsilon$?'' Indeed this problem can be rephrased as ``Does there
exist a path $\gamma$ from $x_i$ to $x_f$ such that for all points
$x \in \gamma$ we have $V(x)-V(x_i) < \epsilon$?'', which (after
some suitable discretization) fits the $\Sig2$ template. In other
words, $\Sig2$ problems are decision versions of two step min-max
problems. While there is no proof that these are more difficult than
either \NP\ or \coNP, one can continue anyways and iterate this
construction, obtaining classes $\Sig{k}$ which answer a question
with a series of $k$ alternating quantifiers.  An example of such a
question would be, given a two-player game (in which both players
have perfect information, and the number of options is finite) and
with a winner after $k$ moves on each side, who has a winning
strategy?  Again, these are clearly finite problems, which would
appear to become more and more difficult with increasing $n$.

The union of such problems defines the ``polynomial hierarchy'' \PH\
of complexity classes (see \cite{Rudich}, and also \cite{Varga} for
a short introduction and a physics study of such problems). Now its
entire definition rests on the premise that $\NP\ne\coNP$, so that
existential quantification is different from universal
quantification.  Conversely, if the two are the same (as would be
the case if $\DP=\NP$ or $\coNP$), this entire hierarchy would
collapse to the simplest case of \NP. While not disproven, this
would lead to all sorts of counterintuitive claims that certain
problems which seem much harder than others actually are not, which
would be very surprising, leading to general acceptance of the
premise $\NP\ne\coNP$.  The general style of argument shows an
amusing resemblance to the generally accepted arguments for
dualities between quantum field theories, string theories and so on,
in theoretical physics (though here the point is the opposite, to
argue that naively similar classes are in fact different).

The upshot of all this is that, while from the point of view of
predictivity the measure factor \eq{ExpLam} is very strong, in
principle determining a unique candidate vacuum, using it
computationally is even more difficult than the \NP\ hard problems
we discussed earlier.

\subsection{Even more difficult problems}

For completeness, and to perhaps clear up some misconceptions, we
should point out that there are even more difficult problems than the
ones we considered.  After \P, a natural next deterministic class to
define is \EXP, the problems which can be solved in time which grows
as an exponential of a polynomial of the problem size.  One could instead
restrict the available space.  For example, \PSPACE\ is the general
class of problems which can be solved with storage space which grows
at most as a polynomial in the problem size.

An easy and possibly relevant inclusion is $\PSPACE\subseteq\EXP$.
Since a computer with $N$ bits of storage space only has $2^N$
distinct states, this is the longest time it could possibly run
without getting caught in a loop (one might call this the Poincar\'e
recurrence time).  Thus, all \PSPACE\ problems can be solved in
finite (though perhaps exponentially long) time.

We also have $\NP\subseteq\PSPACE$ (as are all the classes we
discussed previously).  To show this, we need to show that a program
which generates all candidate solutions only needs polynomial space.
This is easy to see for \SAT, for example.

Of course, once one allows infinite sets into the discussion, one
can have unsolvable problems, such as the Turing halting problem
(decide whether a specified Turing machine halts on every input, or
not).  Unsolvable problems also arise in areas of mathematics which
are closer to physics; perhaps the most relevant for string theory
and quantum gravity is the following
\begin{quotation}
\noindent {\bf Theorem} For no compact $n$-manifold $M$ with $n>4$
is there an algorithm to decide whether another manifold $M'$ is
diffeomorphic to $M$.
\end{quotation}
\noindent (due to S. P. Novikov; see the references and discussion
in \cite{Weinberger}, p. 73).
Here can one imagine $M'$ as
given by some triangulation (of finite but unbounded size), or in
any other concrete way.  This follows by exploiting the
unsolvability of the word problem for fundamental groups in $d> 4$.

It has been argued \cite{GerHart,Nabutovsky} that this
makes simple candidate definitions of the quantum gravity functional
integral, for example as a sum over triangulations of a manifold
$M$, uncomputable even in principle.  While paradoxical, the idea is
not in itself inconsistent; rather it would mean that such a
physical model can in principle realize a larger class of computable
functions than the original Church-Turing thesis.  Indeed, if we
believed in such a model, we might look for ways to make physical
measurements which could extract this information, much as many now
seek to build quantum computers to do computations more quickly than
the classical model of computation allows.

While there is no evidence for this type of uncomputability in
string theory, at present we seem far from having a complete enough
formulation to properly judge this point.  But there are interesting
indirect consequences of these arguments for the structure of the
landscape, as discussed for the geometry of Riemannian manifolds in
\cite{Weinberger,Nabutovsky} and as we intend to discuss for the
string theory landscape elsewhere.

\subsection{Anthropic computing}
\label{subsec:anthropic}

We now take a step back on the complexity ladder. As we mentioned in
section \ref{subsec:ccprob}, one approach to vacuum selection is
environmental selection, also known as the anthropic principle.
Adding this ingredient clearly affects one's expectations of the
ability of cosmological dynamics to ``compute'' vacua with small
cosmological constant or other particular properties. Our detailed
discussion of cosmology appears in \cite{paperII}, but let us review
here what kind of problems one could solve efficiently with
a probabilistic computer when one allows for postselection on part
of the output.

There are precise definitions of a complexity classes which allow
for postselection. For quantum computers this is the class \PostBQP\
(Bounded Quantum Polynomial time with Postselection) recently
defined and studied by Aaronson \cite{Aaronson}. These are the
problems that can be solved in polynomial time using an ``anthropic
quantum computer''. The simplest and most colorful way to describe
such a computer is that we give it an input, let it run, and
postselect on the output satisfying some condition $X$, by killing
ourselves if $X$ is not true. \PostBQP\ is then the class of
problems that can be solved (probabilistically) in polynomial time
by such a machine, \emph{assuming} we survive. The difference with
an ordinary quantum computer is thus that we are allowed to work
with conditional probabilities instead of absolute probabilities.

The analogous class for classical probabilistic computers is
\PostBPP, which turns out to be equal to a class which was defined
before computer scientists started thinking about the power of
postselection, namely $\BPPpath$ \cite{BPPpath,complexityzoo}, and
therefore this is the name usually used for this class. We will
define $\BPPpath$ and explain why it equals \PostBPP\ at the end of
this subsection.

It is easy to see that \PostBQP\ and \PostBPP\ include \NP.  In fact
these classes are larger than \NP, but not unlimited; for example
they are believed to be strictly smaller than \PSPACE\ and \EXP.

The formal definition of \PostBQP\ is as follows. It consists of the
languages $L$ (a language is a particular set of $N$-bit strings for
each $N$) of which membership can be determined as follows:
\begin{itemize}
\item We consider a quantum computer, in other words a unitary
time evolution $U$ acting on some Hilbert space $\cal H$, with $U$
built out of a number of quantum gates (elementary unitary
operations acting on a small number of qubits) which grows
polynomially in the size $N$ of the input. The Hilbert space $\cal
H$ has a tensor product decomposition
$$
{\cal H} \cong {\cal H}_1 \otimes \CV \otimes \CW
$$
where ${\CV}\cong \CW\cong \IC^2$, with basis $\ket{0_\CV}$ and
$\ket{1_\CV}$ (resp. $\CW$).
\item A computation is defined as follows.  We supply an input, a vector
$v\in\cal H$ encoding the string $x$ of which we want to decide
whether it belongs to $L$, and receive an output $Uv$. We insist
that the probability for measuring $\ket{1_\CV}$ in $Uv$ be nonzero
for any input. The output is then the value of a measurement of the
bit $\CW$, conditioned on measuring $\ket{1_\CV}$ in $\CV$.
\item We require probabilistic correctness, meaning that if $x \in L$,
the output is $\ket{1_\CW}$ with conditional probability at least
$2/3$, and that if $x \notin L$, the output is
$\ket{0_\CW}$ with conditional probability at least $2/3$.
\end{itemize}
As in our definition of \BQP\ in section \ref{subsec:quantum}, the
precise number $2/3$ here is not significant as one can achieve a
reliability arbitrarily close to 1 by repeating the computation.

In \cite{Aaronson}, it is proven that $\PostBQP=\PP$, in other words
that the computations which can be performed this way are those in
the class \PP, a probabilistic but classical complexity class.  The
definition of \PP\ is the class of problems which can be ``solved''
by a classical randomized computer (one with access to a random
number source), in the sense that the output must be correct with
probability greater than $1/2$.

This should be contrasted with another class, \BPP, which is the
class of problems which can be solved with probability of
correctness and soundness greater than $2/3$.  While these two
classes may sound similar, they are vastly different, as it is
generally believed that $\BPP=\P$ (and proven that it is contained
in $\Sig2\cap \Pitwo\subseteq \PH$ \cite{Lau83}), while \PP\ is huge.
The point is that, given an error probability $p$ bounded {\bf
strictly} below $1/2$, one can run the same computation many times
to achieve an exponentially small error probability, so \BPP\ is
almost as good as \P\ for many purposes, and much used in real world
computing.

On the other hand, since even flipping a coin has error probability
$1/2$, having an error probability less than $1/2$, but no stricter
bound, is not so impressive.  A computer which produces a correct
output for even the tiniest fraction of inputs, becoming negligible
as the problem size increases, and otherwise flips a coin,
would qualify as \PP.%\footnote{
%And thus the name \PP, which stands for ``particle phenomenologist.''
%Just kidding.}

An example of a computation in \PP, which is believed not to be in
\NP\ is: given a matrix M and integer $k$, is the permanent of M
(defined like the determinant, but with all positive signs) greater
than $k$? Indeed, this problem is \PP-complete, meaning it is not in
\NP\ unless $\NP=\PP$.

Despite its size, the class \PP\ is believed to be smaller than
\PSPACE, not to mention \EXP\ and larger classes.  For example, it
is not even clear that it contains \PH, and there is an oracle
relative to which it does not.

An example of a problem believed not to be in \PP\ is the question
of whether the game of Go has a winning strategy for one of the
players, which (if we allow $n\times n$ boards) is in fact
\PSPACE-complete \cite{Pap94}.  The difference between this and the
simpler game theory problems we mentioned as being in \PH\ is the
length of the game, which is fixed in \PH\ but can depend
(polynomially) on the problem size here.

There are a number of surprising and suggestive equivalences of
postselection classes with superficially different looking classes.

First, as we mentioned already in the beginning, $\PostBPP =
\BPPpath$. The former is defined as the problems solvable on a
probabilistic classical Turing machine in polynomial time with
probability of error less than 1/3, allowing for postselection on
the output. A probabilistic classical Turing machine can also be
thought of as a nondeterministic Turing machine in which 2/3 of the
paths accept if the answer is yes, and 2/3 reject if the answer is
no. In this representation, all computation paths must have equal
length. The probabilistic interpretation is naturally obtained from
this by choosing a random path, with each step choice at a vertex in
the tree having equal probability of being picked (1/2 if the paths
split in two at each step). $\BPPpath$ is similarly defined
\cite{BPPpath,complexityzoo}, but now without postselection, and
instead allowing paths of different length (all polynomial).
Probabilities can still be assigned proportional to the number of
paths accepting or rejecting, but now this is not the same anymore
as assigning stepwise equal probabilities.

That the two classes are equal can be seen as follows. First,
\PostBPP\ is contained in $\BPPpath$. This is because, if we want to
postselect on a property $X$, in $\BPPpath$ we can just create
exponentially many copies of all computation paths for which
property $X$ is satisfied (by continuing the branching process), and
not create copies of the paths where it is not, till the
overwhelming majority of computation paths satisfy property $X$.
This effectively postselects on $X$.

Second, $\BPPpath$ is contained in \PostBPP.  This is because, in
the computation tree of a $\BPPpath$ machine, we can extend the
shorter paths by a suitable number of branchings till all paths have
equal length, labeling all but one of the new paths for each old
path by a 0, and the other paths by a 1. Then in \PostBPP, we can
postselect on paths labeled 1.

Similar equivalences to classes that modify standard probability
rules are true in the quantum case. \PostBQP\ equals the class of
problems that can be solved in polynomial time by a quantum computer
with modified laws of quantum mechanics: either by allowing
non-unitary time evolution (re-normalizing the total probability to
1 at the end), or by changing the measurement probability rule from
$|\psi|^2$ to $|\psi|^p$ with $p \neq 2$ \cite{Aaronson}.

%\rem Do we want to say something about Aaronson's proof here? \rem

%While such extremely difficult problems may not be of relevance for
%cosmology, we find the idea that even postselection does not provide
%infinite computational power to be quite significant. In particular,
%it appears to be a non-trivial claim that, if we can reliably decide
%that our universe is in vacuum $v$ (or predict any other observable
%reliably, meaning establish its likelihood with probability bounded
%strictly above $1/2$), then assuming our universe has the power of
%standard quantum mechanics,
%%(no duality enhancements)
%this computation must be in \PostBQP.
%
%Therefore, any candidate ``selection principles,'' either
%pre-selection, post-selection, or any combination, must be testable
%in \PostBQP, or else the universe is doing something more powerful
%than standard quantum mechanics.

\subsection{Advice} \label{subsec:advice}

Postselection classes quantify the power of future boundary
conditions. What about the power of past boundary conditions? This
is quantified by the notion of \emph{advice}. Classical advice is
extra information $I(N)$ delivered to the Turing machine, depending
only on the input size $N$, and not longer than a prescribed number
$f(N)$ of bits. Thus one defines for example the class \Ppoly, the
set of problems that can be solved in polynomial time with
polynomial length advice. This is believed not to contain \NP. It is
also not contained in \NP\ --- in fact it even contains some
undecidable problems. An example of advice would be a partial list
of solutions of the problem for input length $N$. Note that for any
decision problem, advice of length $2^N$ allows to solve the problem
trivially, since we can simply give a list of all (yes/no) answers
for all possible inputs of length $N$, which has length $2^N$.

Quantum advice is defined similarly, but now the input can be
thought of as some state described by $f(N)$ qubits. For example
\BQPqpoly\ is the class of problems that can be solved by a quantum
computer in polynomial with polynomial length quantum advice. Since
the dimension of the Hilbert space spanned by $f(N)$ qubits is
$2^{f(N)}$ dimensional, and could therefore in principle easily
encode all solutions for all possible inputs, one might be tempted
to conclude that this would be as powerful as exponential classical
advice. However, there are very strong limitations on how much
information can usefully be extracted from a quantum state, and
indeed in \cite{Aaronson:advice} it is shown that $\NP \nsubseteq
\BQPqpoly$ relative to an oracle, and that $\BQPqpoly \subset
\PPpoly$, implying this class is not unlimited in scope. This
supports the picture that an $N$-qubit quantum state is ``more
similar'' to a probability distribution over $N$-bit strings than to
a length $2^N$ string.

This ends our brief tour of complexity theory.  Many of the ideas we
introduced here will find interesting applications in part II.

\section{Practical consequences}
\label{sec:practical}

By ``practical'' we mean the question of how we as physicists trying
to test string theory, or more generally to develop fundamental
physics, should respond to these considerations.  Of course the first
response should be to focus on easy aspects of the problem, and avoid
hard ones.  While at present almost any problem one poses looks hard
to do in generality, we believe there is a lot of scope for clever
algorithms to enlarge the class of easy problems.  But it is valuable
to know beforehand when this is possible, and conversely to realize
when a problem as presently formulated is intractable.

Since getting any handle on the set of candidate string vacua is 
so difficult, in \cite{Douglas:2003um} 
a statistical approach was set out, which has been pursued in
\cite{Ashok:2003gk,Denef:2004ze,Denef:2004cf,Conlon:2004ds,
Blumenhagen:2004xx,Douglas:2005df,Gmeiner:2005vz}
and elsewhere.  A short recent overview is \cite{Kumar:2006tn}.

There is a fairly straightforward response to these issues in a
statistical approach.  It is to make the best use of what information
and ability to compute the observables we do have.  To do this, we
should combine our information into a statistical measure of how
likely we believe each candidate vacuum is to fit the data, including
the cosmological constant other couplings, and discrete data.  This
can be done using standard ideas in statistics; let us outline how
this might be done for the c.c., leaving details for subsequent work.

To try to prevent confusion at the start, we would certainly not
advocate the idea that ``our'' vacuum must be the one which maximizes
such a probability measure, which is clearly as much an expression of
our theoretical ignorance as of the structure of the
problem.\footnote{ The same comment of course applies to the vacuum
counting measures introduced in \cite{Douglas:2003um}.  There, the
theoretical ignorance we were expressing was our lack of knowledge of
what selects a vacuum; indeed a main point made there was that useful
measures can be defined which do not assign probabilities to vacua at
all.}  Given additional assumptions, this might be an appropriate
thing to do, or it might not.  What one should be able to do is
compare relative probabilities of vacua, always making clear the
additional assumptions which entered into defining these.

Thus, we begin by imagining that we have a set of vacua $V$ with index
$i$, in which the cosmological constant is partially computable.
(The same ideas would apply to a larger set of couplings, or other
observables.)
As a simple model, we might consider our set to be a class of string
theory vacua, all of which realize the Standard Model at low energies,
and with a classical (or ``bare'') 
contribution to the cosmological constant modelled by
the BP model.  In other words, the data $i$ specifying a vacuum is a vector
of fluxes, and the classical cosmological constant $\Lambda_{0}$ is given by
the formula \eq{BPmodeldef}.

Thus, our effective field theory is the Standard Model coupled to gravity,
which we regard as defined at the cutoff scale $\mu\equiv 1\, \TeV$.  
The observed value of
the cosmological constant will then be a sum of the classical term and
a series of quantum corrections, both perturbative and non-perturbative,
\begin{equation} \label{eq:vacseries}
\Lambda = \Lambda_{bare} + g^2  F_2(\Lambda)
+ g^4 F_4(\Lambda) + \ldots + e^{-F_{NP}(\Lambda)/g^2} + \ldots .
\end{equation}

The leading quantum correction $F_2$ will be given by a sum of one
loop Feynman diagrams, and depends on all masses of particles, and
other couplings.  As this is an effective field theory, it involves an
integral over a loop momentum $|p|\le \mu$, the cutoff, so it is
finite, but depends on $\mu$ as well.  Finally, $F_2$ depends on the
cosmological constant $\Lambda$ as well, because the graviton
propagator enters in the graviton one loop diagram.  Now, we are most
interested in finding vacua with very small $\Lambda$, and for this
problem we can set $\Lambda=0$ in this propagator and
self-consistently impose $\Lambda=0$ at the end.  Similar arguments
can be made for the higher order terms, and the final result is a
constant shift $\Lambda_{SM}$ to the quantity $\Lambda_0$ defined in
\eq{BPmodeldef}.

Thus, quantum corrections due to the Standard Model do not modify the
previous discussion in any qualitative way.  However, to actually find
the vacua with small $\Lambda$, we must know the quantity $\Lambda_{SM}$
to a precision $10^{-60} \mu^4$ or so.  Since the vacua with small 
$\Lambda$ in the BP model, and all the other landscape models we know of,
are widely spread through configuration space, even a tiny error here will
spoil our ability to pose the problem, even leaving aside the later
complexity considerations.

On general grounds, a perturbative series expansion such as
\eq{vacseries} is an expansion in the marginal couplings
$\alpha_i/2\pi$, where $\alpha_i$ include the gauge couplings
$g_i^2/4\pi$ in each of the three gauge groups, as well as the Yukawa
couplings.  These range from order $1$ for the top quark Yukawa,
through $1/20$ or so for QCD at $1 \TeV$, down to $1/1000$ or so for
the electroweak $U(1)$.  The QCD and top quark contributions are
particularly problematic, as these series are asymptotic with typical
maximal accuracy obtained by truncating the series after about
$1/\alpha$ terms, in other words $\Delta\Lambda \sim
(\alpha/2\pi)^{1/\alpha}$, so the desired accuracy is unattainable.
Solving this problem and doing a reliable
computation requires a non-perturbative framework, such as lattice
gauge theory.  Even before we reach this point, since the number of
diagrams at a given loop order grows factorially, we encounter what
may be intractable computational difficulties.

In contrast to the BP model and the stringy landscape, we will not claim
that we know that this problem is intractable.  It
clearly has a great deal of structure, and we know of no reason
in principle that a clever algorithm could not exist to compute the
single number $\Lambda_{SM}$ (given some precise definition for it) to
arbitrary precision.  On the other hand, it is clearly formidable.
For the foreseeable future, one can only expect precise statements
at leading orders, with hard work required to extend them to each 
subsequent order.

This might not sound like a reasonable physical problem to work on.
We would agree, but nevertheless, let us consider the problem of using
the data at hand, say the first one or two orders of the series 
expansion \eq{vacseries}, along with some lattice gauge theory results,
to improve our estimate of how likely a given vacuum is to describe our
universe.  What we would need to do first is derive a probability 
distribution
$$
P(i,\Lambda)
$$
which expresses the likelihood that vacuum $i$ (in the toy model, $i$
is a list of fluxes), has cosmological constant $\Lambda$.
We start by taking the reliable results, the first orders of 
\eq{vacseries} and the (by assumption) exact data from the BP model,
as the center of the distribution, call this 
$\Lambda_{BP}+\Lambda_{SM~approx}$.
We then need to get an error estimate
for the next order, and make some hypothesis about how this error is likely
to be distributed.  Say this is Gaussian; we come up with the distribution
$$
P(i,\Lambda) = \frac{1}{(2\pi)^{1/2}\sigma}
e^{-(\Lambda-\Lambda_{BP}-\Lambda_{SM~approx})^2/2\sigma^2}
$$
where $\sigma^2\propto g^4$ is the variance, estimated by the size of
the first correction we dropped.

Obviously estimating $\Lambda$ in a real string theory vacuum
would be far more complicated, but the definition of
the single vacuum distributions $P(i,\Lambda)$ should
be clear.  If we can compute them, what should we do with them?
This depends on other assumptions; in particular
the assumption of a prior measure factor on the vacua.  

For definiteness, let us consider our standard $\exp c/\Lambda$
measure factor.  In this case, we need to decide which vacuum realizes
the minimum positive value of $\Lambda$.  Of course, we cannot literally
do this given the data at hand, but what we can do is find a probability
distribution
$$
P_{MIN-CC}(i)
$$
which gives the probability with which the vacuum $i$ would realize
this minimum positive value, if the distributions $P(i,\Lambda)$ were
accurate.

If we strictly follow the definition \eq{vacseries}, the SM contribution
will give a constant shift to all the vacuum energies, so
the energies $\Lambda_i$ of different vacua are highly correlated.
Since the various choices of flux
and configuration will affect the vacuum energy in the hidden sector
as well, this is probably not very accurate; we would suspect that
taking the individual $\Lambda_i$ as independent random variables
is likely to be a better model of the real string theory ensemble.
In any case, let us first assume
independence for simplicity, and then return to the original ensemble.
Since the distributions $P(i,\Lambda)$ are smooth near zero, a good
approximation to $P_{MIN-CC}$ would simply be
\begin{equation}\label{eq:pminone}
P_{MIN-CC}(i) = \frac{P(i,0)}{\sum_j P(j,0)}.
\end{equation}
Naively, the way that vacuum $i$ can achieve the minimum is for it to
realize $\Lambda=\epsilon>0$ for some extremely small value of 
$\epsilon$.  Since the distributions are smooth, we can simply take
$\Lambda=0$ in evaluating the distribution.  There is then a factor for
the expected width of the $\Lambda$ range over which vacuum $i$ really
is the minimum, but given independence
this factor is the same for all vacua, and cancels
out.  We then normalize the resulting distribution to obtain \eq{pminone}.

It is not any harder to do this for the actual assumptions we made
in our previous discussion, according to which there is a constant
shift $\Lambda_{SM}$ to the cosmological constant for all vacua, independent
of the choice of vacuum.  Now it is more efficient to think of the
SM computations as providing a probability distribution 
$$
P_{SM}(\Lambda_{SM}) = \frac{1}{(2\pi)^{1/2}\sigma}
e^{-(\Lambda_{SM}-\Lambda_{SM~approx})^2/2\sigma^2} .
$$
The resulting probability distribution for vacua is that each vacuum
with cosmological constant $\Lambda$ appear with equal weight,
the probability that the SM really did produce the needed shift of the
c.c. to give it a near zero value.  The only difference between this
and the previous discussion is the width factor, which is the
difference between the $i$'th c.c. and the next higher c.c. in the
discretuum,
$$
\Delta\Lambda_{BP}(i) = \Lambda_{BP}(i')-\Lambda_{BP}(i) .
$$
This leads to
\begin{equation}\label{eq:pmintwo}
P_{MIN-CC}(i) = \frac{\Delta\Lambda_{BP}(i)P_{SM}(-\Lambda_{BP}(i))}
{\sum_j \Delta\Lambda_{BP}(j)P_{SM}(-\Lambda_{BP}(j))}
\end{equation}
This is clearly much harder to compute than \eq{pminone}.  In practice,
independence between the c.c.'s in different vacua is probably a more
realistic assumption, and we see that in fact this helps us.

An interesting point about this is that the original 
{\it MIN-CC} distribution did 
not need an additional prior (it was the prior) 
and thus led to a definite prediction
(the minimum c.c.).  By combining this with our theoretical ignorance, we
obtain another probability distribution, again without an explicit prior.
Of course, this is because we are now postulating
a precise model of our theoretical ignorance.

In principle, one could use similar ideas to deal with the difficulty
of actually finding the flux vacua which realize the correct c.c., by
replacing the computational problem of finding the vacua which work,
with that of estimating the distribution of vacua which work.  This is
not very interesting in our toy model, as the choice of flux had no
observable consequences, but might be interesting in a more
complicated and realistic model in which these choices interact.
How practical is this?  While these issues are certainly not the
limiting factor in our present explorations of the landscape, it is 
conceivable that this sort of computability issue might someday arise.

Another application of these ideas might be to justify simplifying our
picture of the landscape.  For example, there is an optimistic
hypothesis, which would remove most of the difficulty associated with
the c.c.  It is that the detailed distribution of c.c.'s is to a very
good approximation independent of the ``interesting'' choices, which
enter into all the other observable quantities.  This seems to us very
plausible as the c.c. receives additive contributions from all
sectors, including hidden sectors which cannot be directly observed.
In this case, while it is important to know how many vacua realize a
c.c. within range of the observed one (after adding the remaining
corrections), we do not need really need to know which specific vacua
match the c.c. to make predictions; indeed this information would not
be very useful.  We would still be in the position of not being able
to literally find the candidate vacua, but this would be more of
philosophical interest.

Such a picture might be used to justify a style of argument discussed
in \cite{Douglas:2004zg} and references there, which could lead to
qualitative predictions if the number of vacua is not too large.  It
is perhaps best described by a hypothetical example, modelled after
\cite{Arkani-Hamed:2004fb,Susskind:2004uv,Douglas:2004qg,Dine:2004ct}.

Suppose we could show that string theory contained
$10^{160}$ vacua with the property $X$ (say that
supersymmetry is observable at upcoming collider experiments), and
which realize all known physics, except for the observed c.c..
Suppose further that they realize a uniform distribution of
cosmological constants; then out of this set we would expect about
$10^{40}$ to also reproduce the observed cosmological constant.
Suppose furthermore that $10^{100}$ vacua with property $\bar X$ work
except for possibly the c.c.; out of this set we only expect the
correct c.c. to come out if an additional $10^{-20}$ fine tuning is
present in one of the vacua which comes close.  Not having any reason
to expect this, and having other vacua which work, we have reasonable
grounds for predicting $X$, in the strong sense that observing $\bar
X$ would be evidence against string theory.

While we cannot presently make such arguments precise, their
ingredients are not totally beyond existing techniques in string
compactification, up to the point where one needs precise results on
the distribution of the c.c. and couplings.  The preceding
``optimistic hypothesis'' would allow bypassing this point.  Do we
believe in it? It seems fairly plausible for Standard Model
observables, but is perhaps less obvious for other properties, for
example the properties of the dark matter.  Rather than simply assume
it, one could use the ideas we just discussed to estimate the
correlation between the cosmological constant and other observables,
and verify or refute this independence hypothesis.

\section{Conclusions} \label{sec:conclusions}

The question of trying to understand the computational complexity of
finding candidate vacua of string theory has two senses, a practical
one concerning our efforts as physicists to find these vacua, and a
more conceptual one of whether early cosmology can be usefully thought
of as having in some sense ``found'' the vacuum we live in, by some
process with a definable computational complexity.  We will
address the second sense of the question in a companion paper \cite{paperII},
building on the discussion in section 5 here
to make precise statements such as
``a cosmological model which can reliably find the vacuum with the
minimum positive cosmological constant is more powerful than a polynomial
time quantum computer with anthropic postselection.''

As to the first sense, we argued that, at least in the various
simplified models now used to describe the landscape containing vacua
of string theory, the problems of finding vacua which agree with known
data are in a class generally believed to be computationally
intractable.  This means that, unless we can exploit some structure
specific to string theory, or unless $\P=\NP$,
we cannot expect to find an
algorithm to do this much faster than an exhaustive search through all
vacua.  Since according to the standard anthropic arguments we need at
least $10^{120}$ vacua to fit the cosmological constant, such an
exhaustive search is clearly infeasible.  Similar statements apply
to many (though not all)
aspects of the problem, such as jointly fitting the various
observed parameters of the Standard Model.

Our strongest statement applies to the Bousso-Polchinski model, which
is so similar to well studied \NP\ complete problems that we could apply
standard arguments.  This is of course a crude approximation to the
real problem in string theory, but the known next steps towards making
this more realistic do not seem to us likely to change the situation.  A
concrete model in which this claim could be tested is the problem
discussed in subsection
\ref{subsec:Ftheory} of finding supersymmetric F-theory
vacua with prescribed $V$.

We considered various ways out, such as the use of approximation
methods, or of measure factors derived from cosmology.  Now one can
sometimes show that approximate solutions to problems are tractable,
and it would be interesting to know if finding a vacuum in the BP model
with cosmological constant $|\Lambda|< c/N_{vac}$ is tractable for
some $c$.  On the other hand, clearly if many vacua fit the existing
data, we then face the possibility that they go on to make different
predictions, and the theory becomes more difficult to test and
falsify.  So, while approximation methods are clearly important, this
sort of ``complementarity'' means that we should be careful what we
wish for.

To illustrate the situation with measure factors, we considered one
popular candidate, the measure $\exp~ c/\Lambda$ depending only on the
cosmological constant.  This overwhelmingly favors the vacuum with the
minimum positive cosmological constant, and from a theoretical point
of view makes as definite a prediction as one could possibly hope for.
But from a computational point of view, it is far more intractable
than the mere problem of finding vacua which fit the data.

One can still hope of course that better understanding or a drastic
reformulation of these problems will change the situation.  It
is important to remember that, if the number of candidate string vacua
is finite, the problem of finding them is strictly speaking {\bf not}
\NP-hard or in any other complexity class, as these are asymptotic
concepts which describe the increase in difficulty as we consider
larger problems in some class.  We are merely reasoning from
properties of the general (or even worst) case in families of
problems, to guess at the difficulty of the specific case of interest.
This type of reasoning is good for producing upper bounds but is not
conclusive.  Of course, it would be very interesting to have concrete
proposals for how the type of difficulty we described could be
avoided.  We might even suggest the converse hope that, if a
proposed solution does not entirely depend on specifics of the string
theory problem, it could lead to new computational models or methods of
general applicability.

Even if these difficulties turn out to be unavoidable, this need not
imply that string theory is not testable.  While at present it is not clear
what experiment might prove decisive, there have been many proposed
tests and even claims of observations
that would pose great difficulties for the theory
(an example, as discussed in \cite{Banks:2001qc}, is
a time-varying fine structure constant; the evidence is reviewed
in \cite{Uzan:2002vq}).
One can certainly imagine
finding direct evidence for or against the theory.
In the near term, experiments to start in 2007 at the Large
Hadron Collider at CERN, continuation of the spectacular progress in
observational cosmology, and perhaps surprises from other directions,
are likely to be crucial.  If positive, such evidence might convince
us that string theory is correct, while the problem of actually
finding the single vacuum configuration which describes our universe
remains intractable.

This raises the possbility that we might someday convince ourselves
that string theory contains candidate vacua which could describe our
universe, but that we will never be able to explicitly characterize
them.  This would put physicists in a strange position, loosely
analogous to that faced by mathematicians after G\"odel's work.  But
it is far too early to debate just what that position might be, and we
repeat that our purpose here is simply to extrapolate the present
evidence in an attempt to make preliminary statements which could
guide future work on these questions.

\vskip 0.3in

{\bf Acknowledgements} \vskip3mm

We acknowledge valuable discussions and correspondence with Scott
Aaronson, Sanjeev Arora, Tom Banks, Raphael Bousso, Wim van Dam,
Michael Freedman, Alan Guth, Jim Hartle, 
Shamit Kachru, Sampath Kannan, Sanjeev Khanna,
Greg Kuperberg, Andrei Linde,
Alex Nabutovsky, Dieter Van Melkebeek, 
Joe Polchinski, Eva Silverstein, Paul Steinhardt, Wati
Taylor, Alex Vilenkin and Stephen Weeks.  
We particularly thank Scott Aaronson 
and Wati Taylor for a critical reading of the manuscript.
MRD would like to thank
John Preskill and Gerald Jay Sussman for helping him follow many of these
topics over the years,  Dave McAllester for introducing 
him to the computational aspects of the SK model,
and Jean Gallier
and Max Mintz for an inspiring visit to the UPenn CIS department.

This research was supported
in part by DOE grant DE-FG02-96ER40959.

\appendix

\section{A simple pseudo-polynomial toy landscape solving algorithm}

\label{app:knapsackalg}

Here we give a simple algorithm for solving the toy landscape
problem of section \ref{sec:fluxNP}: are there any $m \in \{ 0,1
\}^N$ such that
\begin{equation}
 \Lambda_0 - V_{\rm min} \leq \sum_{i=1}^N m_i \Delta V_i \leq \Lambda_0 -
 V_{\rm min} + \epsilon,
\end{equation}
where $\Delta V_i > 0$? We assume the $V_i^\pm$ are known to a
precision of order $\delta$. The problem we are considering is then
only sensible of course if $\delta < \epsilon/N$. Since there is an
order $\delta$ uncertainty anyway, we are allowed to make rounding
errors of order $\delta$ in each term of the sum. Hence, choosing
energy units in which $\delta \equiv 1$, we can round off all
$\Delta V_i$ to their closest integers values (as well as
$\Lambda_0$ and $V_{\rm min}$), and work further over the integers.

Define for $K,s \in \IZ^+$ the Boolean function $Q(K,s)$ to be {\tt
true} iff there is an $m \in \{ 0,1 \}^K$ such that
\begin{equation}
 \sum_{i=1}^K m_i \Delta V_i = s.
\end{equation}
What we are eventually interested in is $Q(N,s)$ for $s$ in the
range $\Lambda_0 - V_{\rm min} \leq s \leq \Lambda_0 - V_{\rm min} +
\epsilon$, or more precisely this interval extended by an amount of
order $N \delta$ on both sides, since $N \delta$ is the maximal
error of the sum. If $Q(N,s) = {\tt false}$  in the entire range of
this extended interval, we know the answer to our question is
negative. If $Q(N,s) = {\tt true}$ for at least one $s$ well inside
the original interval, where `well inside' means more than order $N
\delta$ away from the boundary, we know the answer to our question
is positive. If $Q(N,s)$ happens to be {\tt true} only near (i.e.\
within $N \delta$ of) the boundary of the original interval, we
strictly speaking do not know the answer, since it depends now on
the rounding errors.

The algorithm to compute $Q$ is very simple. Note that trivially
$Q(K,s) = {\tt false}$ if $s < 0$ or $s
> s_{\rm max} \equiv \sum_{i=1}^N \Delta V_i$, so one only needs to compute
an $N \times s_{max}$ matrix. Furthermore, the following recursion
formula computes $Q$:
\begin{equation}
 Q(K,s) = Q(K-1,s) \mbox{~or~} Q(K-1,s - \Delta V_K),
\end{equation}
together with the initial condition $Q(0,s) = {\tt true}$ iff $s=0$.
Thus this algorithm computes the required $Q(N,s)$ in
$$
 O(N s_{max}) = O(N \sum_i \Delta V_i / \delta)
$$
steps, where in the last expression we undid our choice of energy
units $\delta \equiv 1$.

\section{NP-completeness of subset sum}

\label{app:NPCss}

Here we show that the subset sum problem is \NP-complete, by
reducing the 3-\SAT\ problem introduced in section
\ref{subsec:introcompl} to it. The proof is standard, but we give it
here for completeness, and to illustrate the sort of reasoning
commonly used in such proofs. The version we present comes from
\cite{Wegener}.

The general 3-\SAT\ problem has $m$ clauses $c_a$, $a=1,\ldots,m$,
each consisting of the disjunction of 3 boolean variables or their
negation, chosen from a set of $n$ boolean variables $\sigma_i$,
$i=1,\ldots,n$. The question is if an assignment of truth values to
the variables $\sigma_i$ exists such that all clauses $c_a$ are
satisfied. A simple example of a $(m,n)=(3,4)$ problem instance is
the following:
$$
 \exists \, (\sigma_1,\sigma_2,\sigma_3,\sigma_4) : (\sigma_1 \vee \bar{\sigma_2} \vee \sigma_3) \wedge
 (\bar{\sigma_1} \vee \sigma_2 \vee \bar{\sigma_4}) \wedge (\bar{\sigma_1} \vee \bar{\sigma_2} \vee
 \bar{\sigma_3}) \, \, ?
$$
We now give a polynomial reduction from 3-\SAT\ to subset sum. Let
$c_{ai}$ be equal to 1 if $\sigma_i$ appears (un-negated) in clause
$c_a$, and equal to 0 otherwise. Similarly, let $\bar{c}_{ai}$ be 1
if the negation $\bar{\sigma_i}$ appears in $c_a$, and 0 otherwise.
Note that $\sum_i c_{ai} + \bar{c}_{ai} = 3$, because each clause
contains 3 literals. Define $2n + 2m$ integers $\{
x_i,\bar{x}_i,u_a,v_a \}$ in digital representation as follows:
\begin{eqnarray*}
 x_i &=& 10^{i-1} + \sum_{a=1}^m c_{ai} \, 10^{n+a-1} \\
 \bar{x}_i &=& 10^{i-1} + \sum_{a=1}^m \bar{c}_{ai} \, 10^{n+a-1} \\
 u_a = v_a &=& 10^{n+a-1}
\end{eqnarray*}
and an integer
$$
 t =\sum_{i=1}^n 10^{i-1} + \sum_{a=1}^m 3 \cdot 10^{n+a-1}.
$$
For the example given above, this would be (with digits running
down):
$$
\begin{array}{rccccccccccccccccc}
 & & x_1 & x_2 & x_3 & x_4 & \bar{x}_1 & \bar{x}_2 & \bar{x}_3 & \bar{x}_4 & u_1 & u_2 & u_3 &
 v_1 & v_2 & v_3 && t \\
 c_3&& 0 & 0 & 0 & 0 &   1 & 1 & 1 & 0   & 0 & 0 & 1   & 0 & 0 & 1   && 3 \\
 c_2&& 0 & 1 & 0 & 0 &   1 & 0 & 0 & 1   & 0 & 1 & 0   & 0 & 1 & 0   && 3 \\
 c_1&& 1 & 0 & 1 & 0 &   0 & 1 & 0 & 0   & 1 & 0 & 0   & 1 & 0 & 0   && 3 \\
 \sigma_4 && 0 & 0 & 0 & 1 &   0 & 0 & 0 & 1   & 0 & 0 & 0   & 0 & 0 & 0   && 1 \\
 \sigma_3 && 0 & 0 & 1 & 0 &   0 & 0 & 1 & 0   & 0 & 0 & 0   & 0 & 0 & 0   && 1 \\
 \sigma_2 && 0 & 1 & 0 & 0 &   0 & 1 & 0 & 0   & 0 & 0 & 0   & 0 & 0 & 0   && 1 \\
 \sigma_1 && 1 & 0 & 0 & 0 &   1 & 0 & 0 & 0   & 0 & 0 & 0   & 0 & 0 & 0   && 1
\end{array}
$$
We claim that the 3-\SAT\ problem we started from is equivalent to
the subset sum problem
$$
 \exists \, k_i,\bar{k}_i,r_a,s_a \in \{0,1\}: \sum_i k_i x_i + \bar{k}_i \bar{x}_i +
 \sum_a r_a u_a + s_a v_a = t \, \, ?
$$
Indeed, this equation is
$$
 \sum_{i} (k_i + \bar{k}_i) 10^{i-1} + \sum_{a} \biggl[ \biggl( \sum_{i} k_i c_{ai} + \bar{k}_i
 \bar{c}_{ai} \biggr) + r_a + s_a \biggr] 10^{n+a-1} = \sum_{i}
 1 \cdot 10^{i-1} + \sum_{a} 3 \cdot 10^{n+a-1}.
$$
Because each coefficient of the powers of 10 that appear lies
between 0 and 5, this equality has to hold digit by digit. For the
first $n$ digits, this gives
$$
 \forall i: k_i + \bar{k}_i = 1,
$$
%The first set of equations is satisfied iff for each $i$ either
%$k_i=1, \bar{k}_i=0$ or $k_i=0, \bar{k}_i=1$. We associate $\sigma_i
%= {\tt true}$ to the former case and $\sigma_i = {\tt false}$ to the
%latter.
which for any choice of $k_i \in \{0,1 \}$ is satisfied by taking
$\bar{k}_i=1-k_i$. The last $m$ digits then result in
$$
 \forall a:
 \biggr( \sum_i k_i c_{ai} + (1-k_i) \bar{c}_{ai} \biggl) + r_a +
 s_a = 3,
$$
which can be satisfied iff there is a choice of $k_i$ such that
$\forall a: \sum_i k_i c_{ai} + (1-k_i) \bar{c}_{ai} \in \{1,2,3\}$.
This is equivalent to the original 3-\SAT\ problem, with
identification $\sigma_i = {\tt true}$ if $k_i=1$ and $\sigma_i =
{\tt false}$ if $k_i=0$, and with the value of the latter sum equal
to the number of satisfied literals in clause $c_a$.

\section{NP-completeness of 0/1-promise version of subset sum}

\label{app:NPCprom}

Here we show that the 0/1-promise version of subset sum introduced
in section \ref{subsec:compBP} is \NP-complete. We recall that this
problem is the following. The input is a list of $K$ positive
integers $y_a$, $a = 1,\ldots,K$ and a positive integer $s$, with
the promise that for any choice of $m_a \in \IZ^+$
$$
 \sum_a m_a y_a = s \quad \Rightarrow \quad m_a \in \{0,1\}.
$$
The question is if there \emph{exists} a choice of $m_a \in \{0,1\}$
such that $\sum_a m_a y_a = s$. We show this problem is \NP-complete
by polynomially reducing the standard subset sum problem to it.

The input of standard subset sum is a list of $N$
positive\footnote{The restriction to positive integers keeps the
problem NP-complete, as follows e.g.\ directly from the proof in
appendix \ref{app:NPCss}.} integers $x_i$, $i=1,\ldots,N$ and an
integer $t$. The question is if there exist $k_i \in \{0,1 \}$ such
that $\sum_i k_i x_i = t$. We can assume that $0 < t < \sum_i x_i$,
since otherwise the problem is trivial.

We reduce this to the 0/1-promise version as follows. Let
 $$
  u=2 N \sum_i x_i
 $$
and define a list of $2N$ positive integers $\{ y_i ,\bar{y}_i \}$
and an integer $s$ as
\begin{equation} \label{ydef}
  y_i = u^{N+2} + u^i + x_i, \qquad \bar{y}_i = u^{N+2} + u^i,
  \qquad s = N u^{N+2} + \sum_i u^i + t.
\end{equation}
We claim that the subset sum problem we started from is equivalent
to
\begin{equation} \label{promprobred}
  \exists \, k_i,\bar{k}_i \in \{0,1\}: k_i y_i + \bar{k}_i \bar{y}_i =
  s
\end{equation}
and that this is an instance of the promise version of the problem,
i.e.\ for any choice of $k_i,\bar{k}_i \in \IZ^+$:
\begin{equation} \label{promisesat}
  \sum_i k_i y_i + \bar{k}_i \bar{y}_i = s \quad \Rightarrow \quad
  k_i,\bar{k}_i \in \{0,1\}.
\end{equation}
The first claim is easily verified by writing out
(\ref{promprobred}) using (\ref{ydef}):
\begin{equation} \label{promprobred2}
 \sum_i (k_i + \bar{k}_i) u^{N+2} + \sum_i (k_i + \bar{k}_i) u^i +
 \sum_i k_i x_i = N u^{N+2} + \sum_i u^i + t.
\end{equation}
For $k_i,\bar{k}_i \in \{0,1\}$ the coefficients of the powers of
$u$ are guaranteed to be less than $u$, hence the equality has to
hold power by power (``digit by digit'' in base $u$). Then for any
choice of $k_i \in \{0,1\}$, $\bar{k}_i$ is uniquely fixed to
$\bar{k}_i = 1-k_i$, and (\ref{promprobred2}) reduces to $\sum_i k_i
x_i = t$, which is the subset sum problem we started from.

It is slightly less trivial to show that the promise
(\ref{promisesat}) is also satisfied, because \emph{a priori} the
$k_i,\bar{k}_i \in \IZ^+$ are unbounded and therefore we cannot
immediately say that (\ref{promprobred2}) has to be satisfied power
by power. However, it can be seen that (\ref{promprobred2}) has to
be satisfied at the highest power of $u$, i.e.\ $\sum_i k_i +
\bar{k}_i = N$, essentially because any deviation would produce a
left hand side much too large or much too small to match the right
hand side. More precisely:
\begin{itemize}
 \item If $\sum_i k_i + \bar{k}_i \geq N+1$, then $\mbox{LHS} \geq (N+1)
 u^{N+2} > \mbox{RHS}$ because $t < u$, so the
 equality cannot be satisfied.
 \item If $\sum_i k_i + \bar{k}_i \leq N-1$, then
  $\mbox{LHS} \leq (N-1)(u^{N+2}+\sum_i u^i + \sum_i x_i) < (N-1)(u^{N+2} + u^{N+1})
  < (N-1) u^{N+2} + u^{N+2} = N u^{N+2} < \mbox{RHS}$, because $\sum_i x_i <
  u$ and $N-1 < u$.
\end{itemize}
Thus, $\sum_i k_i + \bar{k}_i = N$, and since $N < u$, $t<u$ and
$\sum_i k_i x_i \leq N \sum_i x_i < u$, we have that every
coefficient of the powers of $u$ in (\ref{promprobred2}) is strictly
smaller than $u$, so the equation has to hold power by power. In
particular this implies for the $i$-th power that $k_i + \bar{k}_i =
1$, and therefore $k_i,\bar{k}_i \in \{0,1\}$, which is what we had
to prove.

Finally note that the reduction is indeed polynomial: the number of
bits needed to describe the input of the derived 0/1-promise subset
sum problem is polynomial in the number of bits needed to describe
the input of the standard subset problem we started from (because
the number of bits of $u$ is polynomial), and the number of steps
required to do the reduction is clearly polynomial.

This completes the proof.

\end{document}